\documentclass[aps,prb,letterpaper,superscriptaddress,twocolumn,showpacs,floatfix,10pt]{revtex4-1}

\pdfoutput=1

\usepackage{graphicx}
\usepackage{amsmath}
\usepackage{amstext}
\usepackage{amssymb}
\usepackage{amsfonts}
\usepackage{amsbsy} 
\usepackage{dcolumn}
\usepackage{bm}

\usepackage{color}
\usepackage[colorlinks,citecolor=blue]{hyperref}

\usepackage[bbgreekl]{mathbbol}
\usepackage{amscd}

\renewcommand{\t}[1]{\tilde{#1}} 

\newcommand{\zz}{Z_2^g}
\def\tm{\tilde{m}}
\def\tOmega{\tilde{\Omega}}

\newcommand{\bpm}{\begin{pmatrix}}
\newcommand{\epm}{\end{pmatrix}}
\newcommand{\bmm}{\begin{matrix}}
\newcommand{\emm}{\end{matrix}}
\newcommand{\be}{\begin{equation}}
\newcommand{\ee}{\end{equation}}

\begin{document}

\title{Symmetry fractionalization and anomaly detection in three-dimensional topological phases}
\author{Xie Chen}
\affiliation{Department of Physics and Institute for Quantum Information and Matter, California Institute of Technology, Pasadena, CA 91125, USA}
\author{Michael Hermele}
\affiliation{Department of Physics, University of Colorado, Boulder, Colorado 80309, USA}
\affiliation{Center for Theory of Quantum Matter, University of Colorado, Boulder, Colorado 80309, USA}

\begin{abstract}
In a phase with fractional excitations, topological properties are enriched in the presence of global symmetry.  In particular, fractional excitations can transform under symmetry in a fractionalized manner, resulting in different Symmetry Enriched Topological (SET) phases.  While a good deal is now understood in $2D$ regarding what symmetry fractionalization patterns are possible, the situation in $3D$ is much more open.  A new feature in $3D$ is the existence of loop excitations, so to study $3D$ SET phases, first we need to understand how to properly describe the fractionalized action of symmetry on loops. Using a dimensional reduction procedure, we show that these loop excitations exist as the boundary between two $2D$ SET phases, and the symmetry action is characterized by the corresponding difference in SET orders. Moreover, similar to the $2D$ case, we find that some seemingly possible symmetry fractionalization patterns are actually anomalous and cannot be realized strictly in $3D$. We detect such anomalies using the flux fusion method we introduced previously in $2D$. To illustrate these ideas, we use the $3D$ $Z_2$ gauge theory with $Z_2$ global symmetry as an example, and enumerate and describe the corresponding SET phases. In particular, we find four non-anomalous SET phases and one anomalous SET phase, which we show can be realized as the surface of a $4D$ system with symmetry protected topological order.
\end{abstract}

\maketitle

\section{Introduction}

Fractional excitations in a topological phase are characterized by their fractional statistics when braided around each other. In the presence of a global symmetry, they acquire new topological features. In particular, each fractional excitation can transform under the symmetry in a fractional way. For example, in the $\nu=1/3$ fractional quantum Hall system with $U(1)$ charge conservation symmetry, a single quasi-particle carries $1/3$ of the electron charge while the underlying electrons always have integer charges.\cite{Laughlin1983} This is the phenomenon of symmetry fractionalization (SF). Systems with the same topological order and the same global symmetry can have different SF patterns, resulting in different symmetry enriched topological (SET) phases.\cite{Wen2002,Levin2012,Essin2013,Mesaros2013,Hung2013,Lu2013,Xu2013,Gu2014,Barkeshli14arxiv,Fidkowski}  An interesting question is to understand what SF patterns are possible and where they can be realized.

%

Substantial progress has been made in answering this question for $2D$ topological phases. It was realized that the SF type of a quasiparticle is given by a projective representation of the symmetry.\cite{Wen2002,Wen2003d,Kitaev2006a,Levin2012,Essin2013} Moreover, the projective representations of different quasiparticles should be consistent with their fusion rules. That is, if two quasiparticles can be fused into a third one, their projective representations should combine into that of the third one, up to some linear representation of the symmetry. Following this rule, the whole set of possible SF patterns of a particular topological order can be exhaustively listed.\cite{Essin2013} 

%

However, such a counting is overcomplete. It was realized that some of the SF patterns are anomalous, i.e. they cannot be realized in strictly $2D$ systems.\cite{Vishwanath2013,Wang2013,Burnell2013,Fidkowski2013,Metlitski2013,Bonderson2013,Wang2013a,Chen2013a,Chen2014,Cho2014,Kapustin2014,Kapustin14arxiv,Hermele2015} Various anomaly detection methods have been proposed to identify such SF patterns. The central idea of most methods is based on introducing symmetry fluxes into the system and trying to gauge the global symmetry.\cite{Chen2014,Cho2014,Kapustin2014,Kapustin14arxiv,Barkeshli14arxiv,Hermele2015} If the SF pattern can be realized in a strictly $2D$ model, then the global symmetry can be consistently gauged, resulting in a larger topological theory including the original quasiparticles and the symmetry fluxes and charges. On the other hand, if the SF pattern is anomalous, something goes wrong in the gauging process and the anomaly is revealed. Interestingly, it was found that these anomalous SF patterns can be realized on the surface of $3D$ systems with nontrivial Symmetry Protected Topological (SPT) order in the bulk. In this case, the anomaly exposed in the gauging process on the surface is canceled by one coming from the bulk, resulting in a consistent theory.


%

What about $3D$ topological phases with symmetry? Various topological phases have been found in $3D$ systems, including gauge theories and their twisted versions.\cite{Dijkgraaf1990,Walker2012} What happens when the system also has a global symmetry? This is the question we try to answer in this paper. 

In particular, we address the following two parts of the question:
\begin{enumerate}
\item{How to describe symmetry fractionalization patterns in $3D$?}
\item{How to detect anomalies in the symmetry fractionalization patterns?}
\end{enumerate}

New insights are needed to generalize our understanding from $2D$ to $3D$. First, $3D$ topological phases contain loop like excitations that we refer to as quasi-strings. When describing the symmetry action on these excitations, we must take their extended nature into consideration. Secondly, most of the anomaly detection methods proposed depend on the $2D$ nature of the system and do not generalize in a straight-forward way to $3D$. To identify anomalous SF patterns in $3D$, a new method is needed.

%

We address these issues in this paper. In section \ref{SF}, we discuss how to properly describe SF patterns in $3D$, in particular the nontrivial symmetry action on loop excitations. Our description is based on dimensional reduction to $2D$, in particular on examining differences in $2D$ SET order between regions bounded by dimensionally reduced quasi-strings.  We also relate this description to three-loop braiding processes in $3D$.
Such an understanding enables us to list all possible of SF patterns, although some of them may be anomalous.

In section \ref{anom}, we demonstrate how to use the `flux fusion' method to detect anomalies in $3D$ SF patterns. We introduced the `flux fusion' idea in \onlinecite{Hermele2015} where it was used to identify anomalies in $2D$. This method can be straightforwardly generalized to $3D$ and is used in this paper. We briefly review the basic idea of the method before applying it to $3D$. Throughout our discussion, we use the $3D$ $Z_2$ gauge theory with unitary $Z_2$ symmetry as an illustrative example, which we call for simplicity the $Z_2Z_2$ SET. In particular, we find that there are four non-anomalous SF patterns and one anomalous one in this case. In section \ref{sum}, we summarize our results and discuss open questions. Three appendices contain a more detailed treatment of the dimensional reduction procedure and description of symmetry fractionalization on quasi-strings, accounting for all the fusion and braiding properties characterizing the $2D$ SET orders.

In previous studies, several classes of $3D$ SET phases have been analyzed by focusing on the fractional symmetry representations carried by the quasi-particles in the system. For example, Ref.\onlinecite{Xu2013} classified $3D$ $Z_2$ gauge theories enriched with time reversal symmetry and Ref.\onlinecite{Wang2015a} classified gapless $3D$ $U(1)$ spin liquid with time reversal symmetry. In particular, it was pointed out in \onlinecite{Wang2015a} that certain types of time reversal symmetric $U(1)$ spin liquids are anomalous.

\section{Symmetry fractionalization in $3D$}
\label{SF}

Topological excitations in $3D$ can be either point like quasi-particles or loop like quasi-strings. Symmetry fractionalization on quasi-particles in $3D$ works in the same way as on quasi-particles in $2D$, which we review briefly in section \ref{SF_particle}. Quasi-string excitations exist only in $3D$, not in $2D$. To understand SF patterns in $3D$, the key is to understand how symmetries act on quasi-strings. We discuss this in detail in section \ref{SF_string}. To illustrate our discussion, we use the $Z_2$ gauge theory with unitary $Z_2$ symmetry (the $Z_2Z_2$ SET) as an example. To distinguish the two $Z_2$'s, we label the gauge group as $Z_2^g$ and the global symmetry group as $Z_2^s$. The $3D$ $Z^g_2$ gauge theory has a quasi-particle excitation, which we call the gauge charge $e$, and a quasi-string excitation, which we call the gauge flux loop $m$.  For the $Z^s_2$ global symmetry, we denote the symmetry charge as $Q$ and the symmetry flux loop as $\Omega$.  We enumerate all possible ways the unitary $Z^s_2$ symmetry can fractionalize on the $e$ and $m$ excitations. 

Braiding processes involving three loop excitations will play an important role in our discussion, and we use the following notation for such braiding statistics. The statistics angle for a full braid of two loops $i$, $j$ when they are linked with a base loop $k$ is denoted as $\Phi_{i,j;k}$ (e.g. $\Phi_{m,\Omega;m}$).  We can also consider exchange statistics (\emph{i.e.} a half braid) of two identical loops $i$ linked with a base loop $k$; in this case we denote the statistics angle by  $\Phi_{i;k}$. Sometimes we use the label $i_k$ (e.g. $\Omega_m$) to denote $i$ loops linked with a $k$ base loop. We have suppressed these subscripts in $\Phi_{i,j;k}$ (i.e. we are not writing $\Phi_{i_k,j_k;k}$) for simplicity of notation.

In order to define what we mean by a $Z_2^g$ gauge theory, we need to specify the braiding properties of $e$ and $m$.  These properties have nothing to do with the $Z_2^s$ symmetry, and persist if the symmetry is broken.  We take $e$ to be a boson; there are also $Z_2^g$ gauge theories with fermionic quasi-particles carrying the gauge charge, but we do not consider these theories here.  The $e$ quasi-particle feels $m$ as a $\pi$ flux, so that a statistical phase of $-1$ is acquired when $e$ winds around a $m$ quasi-string.  Finally, for three-loop braiding of $m$ quasi-strings, $\Phi_{m,m;m} = 0$.  Considering exchange statistics of the two $m$ loops linked to a $m$ base loop, there are two consistent possibilities, $\Phi_{m;m} = 0, \pi$.  However, these possibilities are not distinct, as they are related by a natural relabeling of excitations.  We can shift $\Phi_{m;m} \to \Phi_{m;m} + \pi$ by binding $e$ particles to the two $m$ loops linked to the base loop. 
Then, because these loops cannot be shrunk to a point, there is not a natural labeling of $m$ v.s. $em$, and we are free to relabel $m \leftrightarrow em$.

When we apply the dimensional reduction procedure, we will need to discuss $2D$ braiding statistics.  The  statistics angle for a full braid of  point excitations $i$ and $j$ in $2D$ is denoted $\theta_{i,j}$, while we write $\theta_{i}$ for the exchange statistics of two $i$ excitations.

\subsection{Symmetry fractionalization on quasi-particles}
\label{SF_particle}

The local action of symmetry on quasi-particle excitations only needs to satisfy the group multiplication relation up to a phase factor. \cite{Wen2002,Wen2003d,Kitaev2006a,Essin2013} (Here we assume that symmetry does not permute the quasi-particles, which is sufficient to discuss the $Z_2 Z_2$ SET.) We write
\be
U_{a}(g_1)U_{a}(g_2) = e^{i\alpha_{a}(g_1,g_2)} U_{a}(g_1g_2) \text{,}
\ee 
where $U_a(g)$ gives the action of symmetry operation $g$ on a quasiparticle of type $a$.


The phase angle $\alpha_{a}(g_1,g_2)$ is not arbitrary. If $n$ copies of $a$ fuse into the vacuum, then $e^{i\alpha_{a}(g_1,g_2)}$ has to be an $n$th root of unity. On the other hand, we can redefine $U_a(g) \to \lambda_a(g) U_a(g)$ for $\lambda_a(g)$ a $n$th root of unity, and any two sets of $\alpha$ related in this way should be considered equivalent. Therefore, $U_{a}(g)$ forms a projective representation of the symmetry group $G$ with coefficients in $Z_n$ and $e^{i\alpha_{a}(g_1,g_2)}$ specifies an element in $H^2(G,Z_n)$. Moreover, if $a$ and $b$ fuse into $c$, then $\{U_{a}(g) \otimes U_{b}(g),g\in G\}$ should be equivalent to $\{U_{c}(g),g\in G\}$, in the sense that both representations are characterized by the same element of $H^2(G, Z_n)$.

Combining these properties together, for unitary internal symmetries with finite symmetry group, it was realized that the symmetry fractionalization pattern in $2D$ is encoded in the projective fusion rules of symmetry fluxes,\cite{Chen2014,Barkeshli14arxiv,Fidkowski,Etingof2010} namely
\begin{equation}
\Omega_{g_1} \Omega_{g_2} = \omega(g_1, g_2) \Omega_{g_1 g_2} \text{,}
\end{equation}
where $\Omega_g$ labels the symmetry flux for a group element $g$, and $\omega(g_1, g_2) \in {\cal A}$, the set of Abelian quasi-particles in the theory.  These fusion rules specify an element of $H^2(G, {\cal A})$.  Because the local action of a symmetry operation $g$ on $a$ is given by a braid between $a$ and $\Omega_g$,\cite{Chen2014} the projective phase factors $e^{i\alpha_{a}(g_1,g_2)}$ are given by the phase factor resulting from a full braid between $a$ and $\omega(g_1,g_2)$.


In the $3D$ $Z_2Z_2$ SET, the quasi-particle $e$ can transform projectively under the $Z_2^s$ symmetry. In particular, there are two possibilities: $e$ carrying integer $Z_2^s$ charge and $e$ carrying half odd integer $Z_2^s$ charge, which we label as $e0$ and $eC$, respectively.  Applying the non-trivial symmetry operation twice to $e$ results in a phase factor of $+1$ for $e0$ and $-1$ for $eC$.   In both cases, two $e$ particles together always carry integer $Z_2^s$ charge, which must occur because they fuse into the trivial sector.

\subsection{Symmetry fractionalization on quasi-strings}
\label{SF_string}

How can symmetry fractionalize on quasi-string excitations? We try to answer this question in this section. First, we discuss the possibility of quasi-string excitations carrying fractional representations, or gapless modes protected by the symmetry. While these are possible nontrivial ways quasi-strings can transform under symmetry, a more complete perspective is provided by viewing quasi-strings as boundaries between $2D$ SET phases upon dimensional reduction to $2D$. We explain this point in detail and then count all possible symmetry fractionalization patterns in the $Z_2Z_2$ SET.

\subsubsection{Quasi-string carrying fractional symmetry representations}

One possible way for quasi-strings to transform nontrivially under symmetry is to carry fractional symmetry representations, just like quasi-particles. However, this is not an intrinsic feature of quasi-strings. 

First, we consider a loop of quasi-string that is not linked with any other loops.  Such a quasi-string can be shrunk down to a point, and the loop becomes a point excitation while any fractional symmetry representation it carries remains unchanged. That is to say, there is some quasi-particle excitation in the theory that carries the same fractional representation. Then, by attaching the anti-particle to the quasi-string, we can cancel the fractional symmetry representation carried by the quasi-string. Therefore, fractional symmetry representation carried by quasi-strings can always be removed, and, as we shall do in the following discussion, we are free to focus on quasi-strings carrying no fractional symmetry representation without any loss of generality.

In fact, unlike quasi-particles, some quasi-string excitations can appear on their own, not in pairs. For example, this is the case for the $m$ flux loop in the $Z_2^g$ gauge theory, although the composite of $em$ has to come in pairs. Quasi-string excitations that can exist on their own cannot carry fractional symmetry representations because they can shrink down to nothing. Quasi-string excitations that come in pairs can carry fractional representations, but that reduces to the fractional representation carried by quasi-particles once the strings are shrunk down to a point.

Quasi-strings that are linked to other loops cannot be shrunk down to a point, so it is not clear, at this point, how to define the fractional symmetry representation they carry.

\subsubsection{Quasi-string carrying gapless modes protected by symmetry}
\label{mSPT}

Quasi-strings are one dimensional excitations. A more intrinsic way for them to transform under symmetry is to carry $1D$ gapless modes protected by the symmetry. Such $1D$ gapless modes appear on the edge of $2D$ symmetry protected topological (SPT) phases. That is to say, we can imagine the quasi-string excitation as bounding a $2D$ surface to which a $2D$ SPT state is attached. This is natural, because quasi-strings in $3D$ gauge theories can be viewed as edges of strongly fluctuating surfaces with vanishing surface tension.

\begin{figure}[htbp]
\begin{center}
\includegraphics[width=7.0cm]{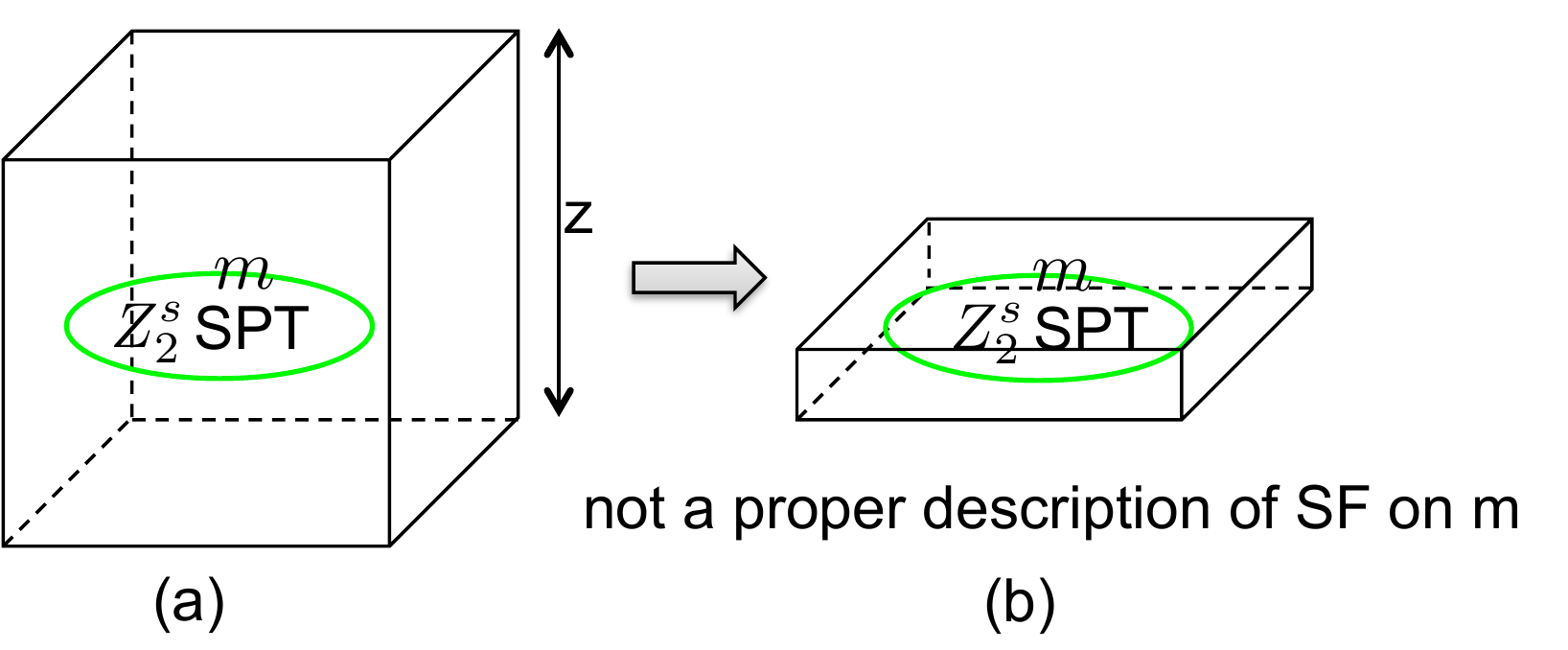}
\caption{One possible way for $Z_2^s$ symmetry to transform nontrivially on $Z_2^g$ flux loop $m$: (a) $m$ loop carries gapless modes protected by $Z_2^s$ symmetry and bounds a $2D$ $Z_2^s$ SPT state; (b) Compressing the dimension of the system perpendicular to the surface bounded by the $m$ loop reduces the system to $2D$. However, depicting the $m$ loop as the boundary of a $2D$ SPT state is an incomplete and inaccurate description of the symmetry fractionalization pattern on $m$.}
\label{mZ2SPT}
\end{center}
\end{figure}

For example, in our $Z_2Z_2$ SET example,  we can have the $m$ loop transform as the boundary of a $2D$ $Z_2^s$ SPT state. If we create an $m$ loop in the bulk of the system, as shown in Fig.\ref{mZ2SPT} (a), naively we would expect it to be gapless unless the $Z_2^s$ symmetry is broken.

In an attempt to examine such a symmetry action in more detail, it is natural to consider a dimensional reduction procedure where we compress the $z$ dimension of the system while keeping the other two dimensions infinite, as shown in Fig.\ref{mZ2SPT} (b). It is tempting to conclude that if an $m$ loop is inserted in the $xy$ plane, then the region inside the string is in the nontrivial $2D$ $Z_2$ SPT phase, while the outside is in the trivial phase. However, this is not an accurate description of the dimensionally reduced system, and thus does not provide a description of symmetry fractionalization on quasi-strings. In order to have a better understanding, we need to look at the dimensional reduction process in a more careful way. This is similar to the dimensional reduction approach used in Ref.\onlinecite{Wang2014a,Wang2015, Moradi2015} to study $3D$ topological phases but here we add symmetry to the discussion.

\subsubsection{Quasi-string as boundary between 2D SET phases}
\label{mSET}

A more careful analysis of the dimensional reduction procedure illustrated in Fig.\ref{mZ2SPT} (b) shows that a proper description of symmetry fractionalization on quasi-strings is obtained by viewing them as boundaries between $2D$ SET phases, and not SPT phases. That is, it is important to take into account the nontrivial topological order of the dimensionally reduced system.  

\begin{figure}[htbp]
\begin{center}
\includegraphics[width=7.0cm]{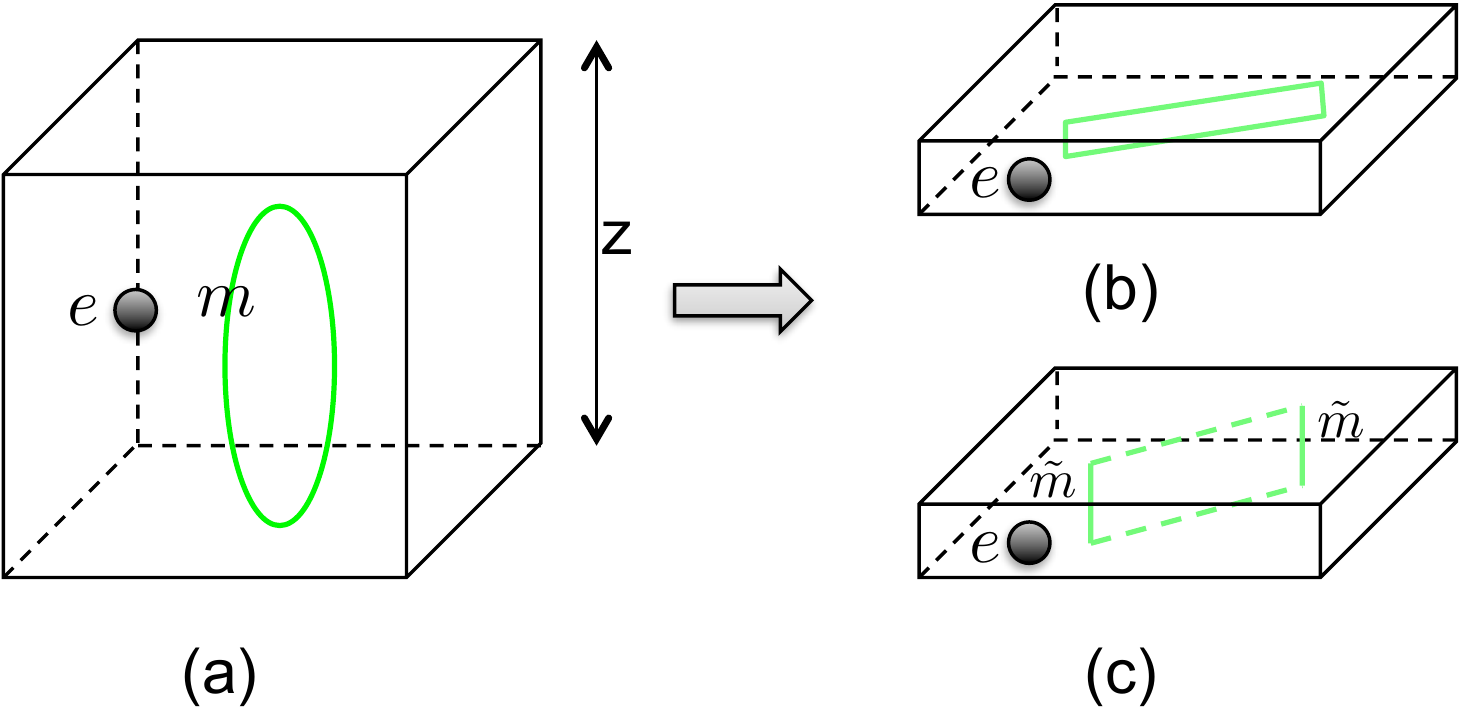}
\caption{dimensional reduction of a $3D$ $Z_2^g$ gauge theory into a $2D$ $Z_2^g$ gauge theory: (a) a $3D$ $Z_2^g$ gauge theory with gauge charge $e$ and gauge flux $m$. After compressing the system in the $z$ direction, the gauge charge remains a quasi-particle while (b) gauge flux loop with finite extent in $z$ direction becomes non-topological (c) a gauge flux loop which extends across the $z$ direction becomes two quasiparticles, which are the gauge fluxes in $2D$.}
\label{DR}
\end{center}
\end{figure}

The first step is to understand what topological order the dimensionally reduced system has, and for this purpose we can temporarily ignore the $Z_2^s$ symmetry. Suppose that we start from a $3D$ $Z_2^g$ gauge theory and compress the system down in the $z$ direction, as shown in Fig.\ref{DR}. We assume periodic boundary conditions in all three directions. The height of the system in the $z$ direction is finite, but larger than any correlation lengths, while the extent in the $x$ and $y$ directions is infinite.


In this geometry, the system becomes a $2D$ $Z_2^g$ gauge theory. To see this, note that the gauge charge $e$ quasi-particles in the $3D$ bulk become quasi-particle excitations in the $2D$ bulk, which are free to move in the $xy$ plane. The other type of $2D$ quasi-particles are $m$ quasi-strings that wind once across the system in the finite $z$ direction; these excitations are quasi-particles in the dimensionally reduced system, rather than quasi-strings, due to their finite extent and, thus, finite energy cost.  In the $2D$ theory, an $e$ excitation going around one of these quasi-particles is equivalent to an $e$ particle going around a $m$ quasi-string in the original $3D$ bulk, which results in a $-1$ phase factor. Therefore, the new quasi-particles correspond to the gauge fluxes in the $2D$ $Z_2^g$ gauge theory and we label them by $\tilde{m}$. The full topological order of the $2D$ system is that of a $Z_2^g$ gauge theory.

Flux loops of extent less than the system size in the $z$ direction, as shown in Fig.\ref{DR} (b), become non-topological excitations in $2D$. For an $e$ quasi-particle to braid around a segment of such a loop, it has to pass a finite distance from the loop during the braiding process. Such a process can be perturbed by various local perturbations, and there is no well-defined statistical phase.  However, starting with such a non-topological flux loop, we can create two $\tilde{m}$ quasi-particles in the $2D$ theory by stretching the loop in the $z$ direction until its top and bottom segments meet and annihilate, as shown in Fig.\ref{DR} (c).

We want to note that there is an ambiguity in what $2D$ $Z_2^g$ gauge theory we can get from this dimensional reduction process. There are two different $Z_2^g$ gauge theories in $2D$, one with bosonic gauge flux ($\tilde{m}$) and the other with semionic gauge flux ($\tilde{m}$).\cite{Dijkgraaf1990} They are called the Toric Code and the double semion topological order respectively\cite{Kitaev2003,Levin2005}. Exactly which one we obtain depends on the details of the dimensional reduction process. While both are possible, this distinction is not important in our discussion, as we will see below. We note that a more detailed discussion of the dimensional reduction procedure for $3D$ $Z_2^g$ gauge theory is given in Appendix A.

\begin{figure}[htbp]
\begin{center}
\includegraphics[width=7.0cm]{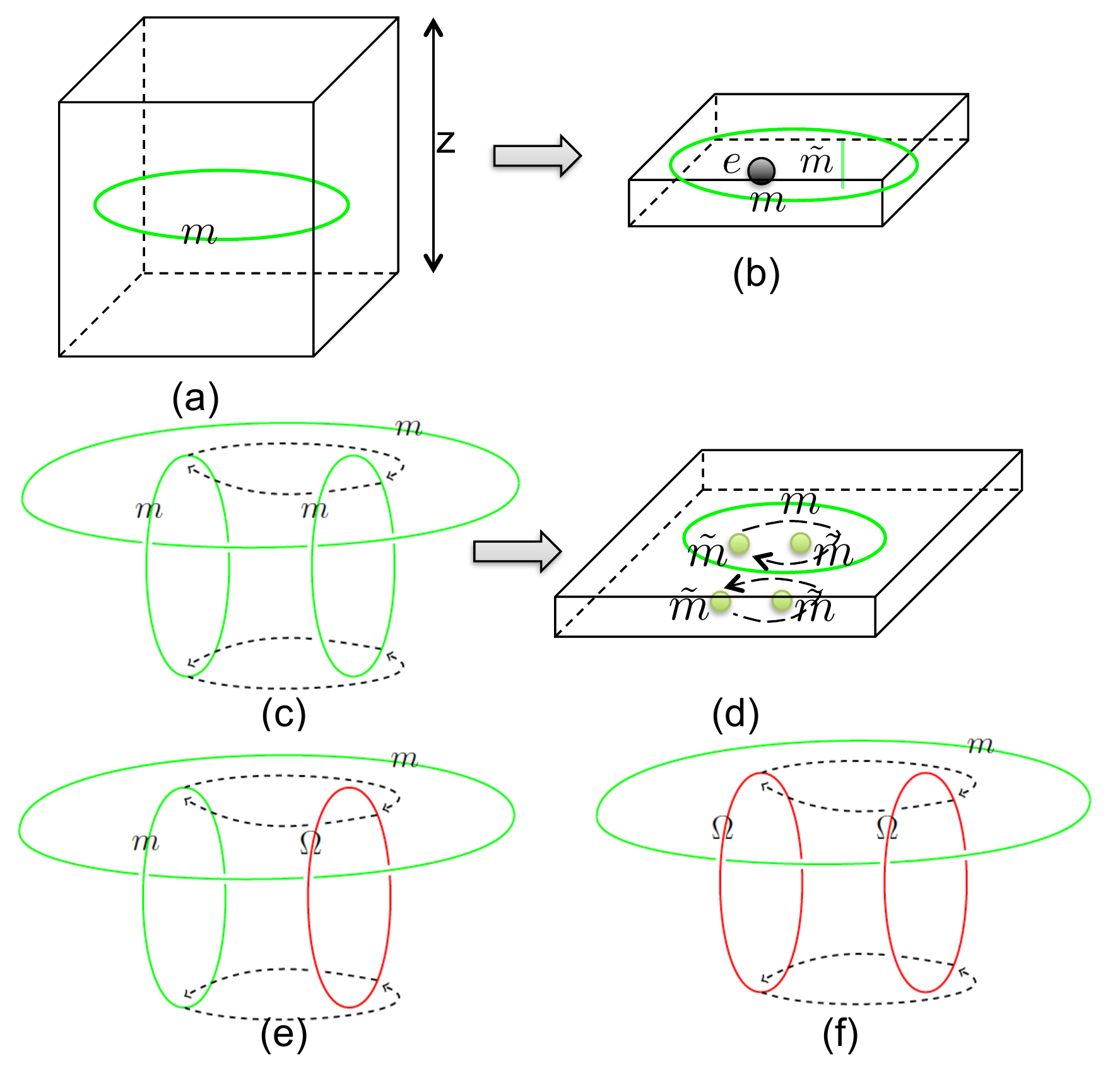}
\caption{Dimensional reduction with an $m$ loop in the $xy$ plane: (a) insert an $m$ loop in the $xy$ plane and (b) compress the system down in the $z$ direction. The resulting $2D$ system has both topological order and symmetry and may carry different SET order inside and outside the $m$ loop. The type of $Z_2$ gauge theory (Toric code or double semion) on the two sides must be the same as (d) the braiding statistics between pairs of $\tilde{m}$ inside and outside of $m$ should cancel which corresponds to (c) the bosonic braiding statistics between two $m$ loops when linked with the $m$ base loop. However, $\tilde{m}$ can carry different symmetry charges which can be detected with the three loop braiding process shown in (e). Also $\t{\Omega}$ may have different topological spin on the two sides of $m$ which can be detected with the three loop braiding process shown in (f).}
\label{DRm}
\end{center}
\end{figure}

Now imagine inserting an $m$ flux loop in the $xy$ plane before compressing the $3D$ system down to $2D$, as shown in Fig.\ref{DRm} (a) and (b). Using the same argument as before, we can show that the resulting $2D$ system is again a $2D$ $Z_2^g$ gauge theory. Moreover, the dimensional reduction process can be done without breaking the $Z_2^s$ symmetry of the system. Therefore, we obtain a $2D$ system with $Z_2^g$ gauge theory type topological order and $Z_2^s$ symmetry. Then we can ask what type of symmetry enriched topological order does it have. The interesting possibility is that, the inside and the outside of the $m$ loop can have different SET orders. Therefore, a proper description of the symmetry fractionalization pattern on $m$ should contain such differences. 

First, we show that the two sides of the $m$ loop must have the same type of $2D$ $Z_2^g$ gauge theory (Toric Code or double semion). That is, $\tilde{m}$ is either both bosonic or both semionic on the two sides. Because two $\tilde{m}$'s fuse to vacuum in the $2D$ theory, the exchange statistics $\theta^{[k]}_{\tilde{m}}$ must be an integer multiple of $\pi/2$.  The superscript $[k] = [I], [m]$ refers to the inside and outside of the base loop, respectively, allowing for the possibility that properties of the $2D$ theory in these regions may be different.  Then, noting that $e \tilde{m}$ is a gauge flux with statistics $\theta^{[k]}_{e \tilde{m}} = \theta^{[k]}_{\tilde{m}} + \pi$, if needed we can redefine $\tilde{m} \to e \tilde{m}$, so that $\theta^{[I]}_{\tilde{m}}, \theta^{[m]}_{\tilde{m}} = 0, \pi/2$.

In fact, $\theta^{[I]}_{\tilde{m}} = \theta^{[m]}_{\tilde{m}}$.  To see this, imagine braiding two pairs of $\tilde{m}$'s, one inside of the $m$ base loop in a clockwise way and one outside of the base loop in a counterclockwise way, as shown in Fig.\ref{DRm}(d). This results in braiding statistics summing to $-\theta^{[m]}_{\tilde{m},\tilde{m}} + \theta^{[I]}_{\tilde{m},\tilde{m}}$. 
Back to the $3D$ system before dimensional reduction, such a braiding process corresponds to the braiding of two $m$ loops when they are linked with a $m$ base loop, with braiding statistics $\Phi_{m,m;m} = 0$. As shown in Fig.\ref{DRm} (d), the loop on the left expands and passes the right side loop from the outside and then shrinks and goes back to its original position through the inside of the right side loop. Throughout the process, the two $m$ loops are both linked with the $m$ loop. This is the three loop braiding process described in \onlinecite{Wang2014a,Jiang2014,Wang2015}. The $\tilde{m}$ fluxes after dimensional reduction can be thought of as the intersection point of the vertical $m$ loops with fictitious surfaces spanning the inside and outside of the base loop. Therefore,
\be
\Phi_{m,m;m} =  \theta^{[I]}_{\tilde{m},\tilde{m}} -\theta^{[m]}_{\tilde{m},\tilde{m}} = 0 \text{,}
\ee
which, combined with the discussion above, implies $\tilde{m}$ has the same exchange statistics on both sides of the $m$ base loop.  Whether $\tilde{m}$ is a boson or a semion depends on the details of the dimensional reduction procedure and is not an intrinsic property of the $3D$ SET phase.

Next, we can ask how the $Z_2^s$ symmetry enriches the dimensionally reduced $2D$ $Z_2^g$ gauge theory. This enrichment can be different on the two sides of the base loop $m$. In particular, $\tilde{m}$ can carry different $Z_2^s$ symmetry charge in the two regions, and also the $Z_2^s$ symmetry flux $\tilde{\Omega}$ can have different exchange statistics. 

Similar to our discussion above, such differences in the dimensionally reduced SET orders are reflected in nontrivial three loop braiding processes back in the $3D$ bulk. 
The difference in the $2D$ exchange statistics of $\tilde{\Omega}$ corresponds to the three loop braiding process shown in Fig.\ref{DRm} (e) with $m$ as the base loop and the two $\Omega_m$ loops braiding around each other. Because $Q \tilde{\Omega}$ is another $Z_2^s$ symmetry flux with exchange statistics shifted from that of $\tilde{\Omega}$ by $\pi$, $\theta^{[k]}_{\tilde{\Omega}}$ is only well-defined modulo $\pi$.  There are thus two distinct possibilities: the exchange statistics of $\tilde{\Omega}$ on the two sides of $m$ may be the same or differ by $\pi/2$.  We label these two cases by $\Omega_m b$ and $\Omega_m s$, respectively, and the related three loop braiding statistics in $3D$ is
\be
\Phi_{\Omega;m} = \theta^{[I]}_{\tilde{\Omega}}-\theta^{[m]}_{\tilde{\Omega}} = 0 \text{\ or } \pi/2 \text{.}
\ee
Because loops linked with a base loop $m$ can be braided and exchanged like quasi-particles in $2D$, we can also describe these two cases in terms of the topological spin of the loop $\Omega_m$.  Using this language,  $\Omega_m b$ ($\Omega_m s$) corresponds to topological spin $\pm 1$ ($\pm i$).

Similarly, the difference in the $Z_2^s$ symmetry charge carried by $\tilde{m}$ in the $2D$ theory is reflected in the braiding process  shown in Fig.~\ref{DRm} (f). There, $\Omega_m$ and $m_m$ loops are linked to a base loop $m$, and $\Omega_m$ is braided around $m_m$.   This is related to the braiding statistics in the $2D$ theory by
\be
\Phi_{\Omega,m;m} = \theta^{[I]}_{\tilde{m},\tilde{\Omega}}-\theta^{[m]}_{\tilde{m},\tilde{\Omega}} \text{.}
\ee
We note that if $\tilde{m}$ carries integer (half-integer) charge, then the statistics angle $\theta_{\tilde{m}} = 0,\pi$ ($\theta_{\tilde{m}} = \pm \pi / 2$).  Therefore, if $\tilde{m}$ inside the $m$ loop carries half $Z_2^s$ charge while $\tilde{m}$ outside the $m$ loop carries integer $Z_2^s$ charge (or vice versa), the corresponding three-loop statistics is $\Phi_{\Omega,m;m} = \pm \pi/2$.  We label this kind of symmetry fractionalization pattern on the $m$ base loop as $m_m C$. Alternatively, $\tilde{m}$ can carry the same charge (integer or half-integer) on both sides of $m$, a symmetry fractionalization pattern we label by $m_m 0$. We can also view braiding $\Omega_m$ around $m_m$ as a way to detect the $Z_2^s$ symmetry charge of the loop $m_m$.  Therefore, we say $m_m 0$ ($m_m C$) corresponds to integer (half-integer) charge of the $m_m$ loop.

A careful analysis of all possible $2D$ SET phases resulting from the dimensional reduction process is given in Appendix B. 

Now we can see why describing the quasi-string as the boundary between two $2D$ SPT phases, as we did in section \ref{mSPT}, is inaccurate. First of all, this is an incomplete description because it does not specify, for example, if $\tilde{m}$ carries different fractional $Z_2^s$ charges on the two sides or not. Moreover, sometimes it is not even a well-defined description, because adding a $2D$ SPT phase on top of a $2D$ SET phase (with the same symmetry) may not change the SET order at all, as observed in Ref.\onlinecite{Lu2013}. Therefore, the difference in $2D$ SPT order on the two sides of the quasi-string cannot be unambiguously defined in the presence of the $2D$ SET order in the dimensionally reduced system. In fact, depending on the symmetry fractionalization pattern, gapless modes carried by the quasi-string may not be stable, and may be removed via interaction with the fractional excitations in the SET phase. 

From the previous analysis, we arrive at the following observation: \emph{A description of symmetry fractionalization on quasi-strings can be obtained via dimensional reduction, where quasi-strings become boundaries between two $2D$ SET phases. The difference in the two $2D$ SET orders describes the symmetry action on the quasi-string.} 

\begin{figure}[htbp]
\begin{center}
\includegraphics[width=7.0cm]{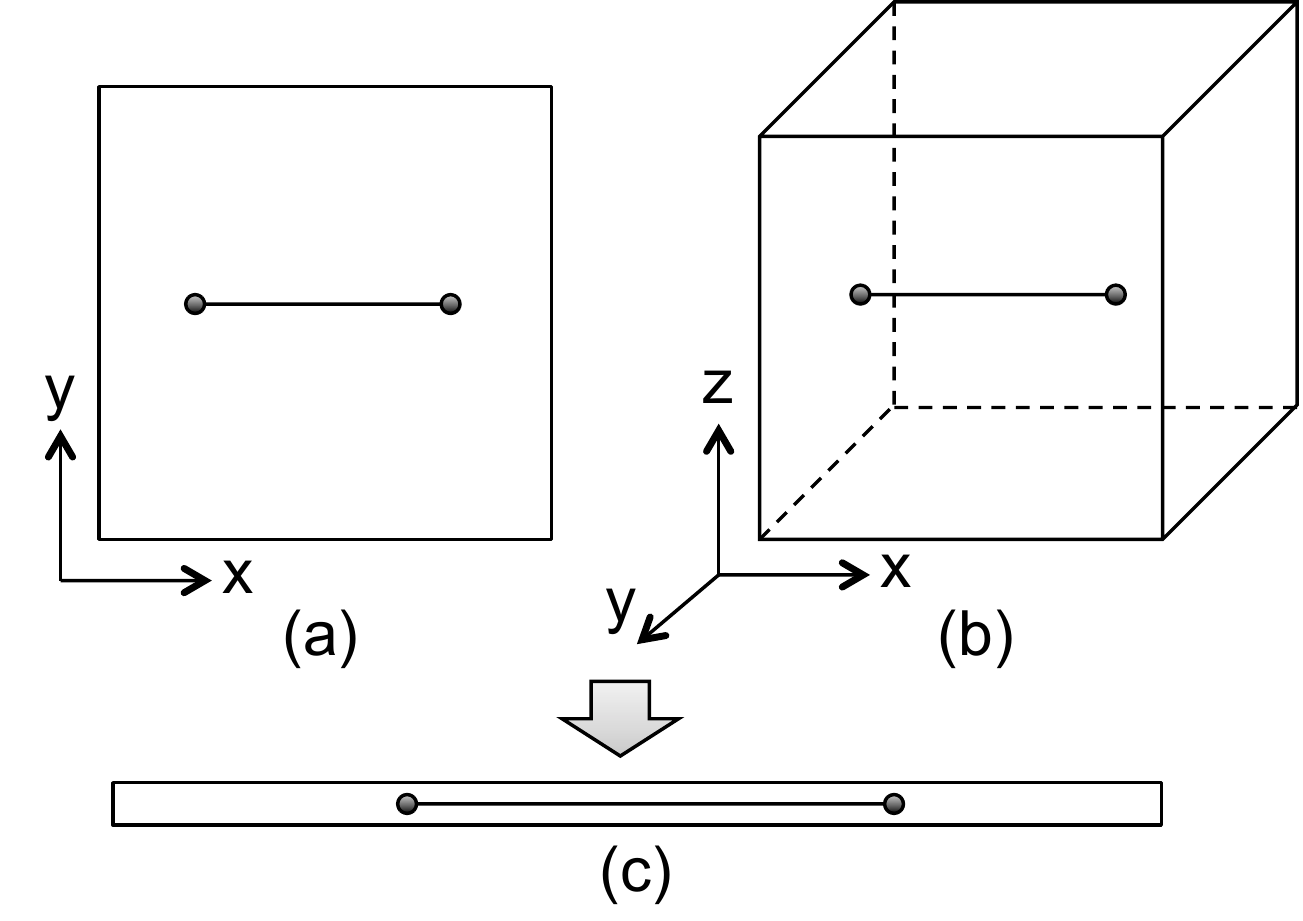}
\caption{dimensional reduction for quasi-particles: create a pair of quasi-particles and pull them apart in the $x$ direction in (a) a $2D$ system or (b) a $3D$ system. Compress the system down in $y$ or $yz$ direction and (c) reduce it to a $1D$ chain. The middle part (between the quasi-particles) of the system can have different SPT order than the outer part and the quasi-particle exists as a boundary state between the two.}
\label{DRe}
\end{center}
\end{figure}

We would like to comment that a similar dimensional reduction procedure can be used to study symmetry fractionalization on quasi-particles, but there are important differences between the quasi-particle case and the quasi-string case. Imagine creating a pair of quasi-particles and pulling them apart in the $x$ direction. This can be done either in a $2D$ SET phase, as shown in Fig.\ref{DRe} (a), or in a $3D$ SET phase [Fig.\ref{DRe} (b)].  We can compress the other dimensions, \emph{i.e.} the $y$ dimension in the $2D$ case and the $y, z$ dimensions in the $3D$ case, and reduce the system down to $1D$. As long as the global symmetry is preserved in the reduction process, we obtain a $1D$ gapped system with symmetry and we can ask what type of symmetric phase it is in. Fractional quasi-particle and quasi-string statistics do not survive dimensional reduction to $1D$, but $1D$ gapped systems with symmetry can have symmetry protected topological (SPT) order. In our dimensionally reduced system, the middle part of the system  (between the quasi-particles) can have different SPT order than the outer part, and the quasi-particles become boundary states between these two regions, as shown in Fig.\ref{DRe} (c).

Based on this discussion, the symmetry fractionalization on a quasi-particle in a SET phase can be characterized by the projective representation of the symmetry carried by the corresponding boundary state between two $1D$ SPT orders.  However, this characterization is not complete, and some information about symmetry fractionalization is lost upon dimensional reduction to $1D$.  This raises the possibility that dimensional reduction to 2D may also give an incomplete description of symmetry fractionalization on quasi-strings; this point is discussed further in Sec.~\ref{sum}.

A major difference from the quasi-string case is that, after the above dimensional reduction for quasi-particles we obtain a $1D$ system, which cannot have fractional excitations or long-range entanglement. Therefore, the dimensionally reduced $1D$ system can only have SPT order, and the quasi-particle becomes a boundary state between SPT orders. On the other hand, after dimensional reduction for quasi-strings, we obtain a long-range entangled $2D$ system that supports fractional excitations. Therefore, the dimensionally reduced $2D$ system has SET order, and the quasi-string exists as a boundary between two $2D$ SET orders.

\subsection{$Z_2^s$ symmetry fractionalization in $3D$ $Z_2^g$ gauge theory}
\label{sym_frac}


Combining the descriptions of symmetry fractionalization on quasi-particles and quasi-strings, we can try to list all possible symmetry fractionalization (SF) patterns in $3D$ SET phases. Naively, one might expect that the total SF pattern can be described by independently choosing the SF pattern for the quasi-particles and the SF pattern for the quasi-strings. However, this turns out to be too simplistic. The SF patterns for quasi-particles and quasi-strings need to be consistent with each other.  Moreover, there may be redundancy in the naive counting, as two seemingly different SF patterns may actually be the same. The general principles are:
\begin{enumerate}
\item{Consistency: The SF pattern of quasi-particles
should be the same both in the 3D  bulk and in
the dimensionally reduced systems used for the description of the SF pattern on quasi-strings.}
\item{Redundancy: Two SF patterns are the same if they
can be related by two kinds of operations:
(a) Redefining symmetry fluxes
when they are linked with nontrivial base loops
by attaching quasi-particles to them. Such quasiparticle
attachment can be different for different
base loops. (b) Redefining symmetry fluxes by attaching
quasi-strings to them. Such quasi-string
attachment should be independent of the base loop.}
\end{enumerate}
In this section, we explain and apply these principles to describe all SF patterns of the $Z_2 Z_2$ SET example. A more detailed discussion, taking into account the full description of the dimensionally reduced SET orders given in Appendix B, is presented in Appendix C.

As we discussed in section \ref{SF_particle}, the gauge charge $e$ can carry either integer or half integer $Z_2^s$ symmetry charge. We label these two cases as $e0$ and $eC$. On the other hand, the SF pattern on the gauge flux $m$ is described as the difference between two $2D$ SET phases with $Z_2^g$ topological order and $Z_2^s$ symmetry. Possible types of $2D$ SET phases with $Z_2^g$ topological order and $Z_2^s$ symmetry have been classified in \onlinecite{Lu2013}. One might want to pick any two possibilities from \onlinecite{Lu2013}, combine it with either $e0$ or $eC$ and produce a total SF pattern. However, the situation is more complicated.

First, the symmetry charge carried by $e$ should be the same whether in the $3D$ bulk or in the dimensionally reduced $2D$ theory. Therefore, once we have chosen $e0$ or $eC$ in the $3D$ bulk, we need to make the same choice in the $2D$ SET phases used to describe the SF pattern on $m$.  We proceed by discussing the cases of $e0$ and $eC$ in turn.

Consider the $e0$ case, \emph{i.e.} we suppose that $e$ carries integer $Z_2^s$ charge.  Therefore, in the $2D$ SET phases resulting from dimensional reduction, $e$ carries integer $Z_2^s$ charge both inside and outside of $m$ (Fig.\ref{DRm}). Now to completely specify the SF pattern on $m$, we need to consider the difference in the $Z_2^s$ charge carried by $\tilde{m}$ inside and outside of $m$, and the difference in topological spin of the $Z_2^s$ flux $\tilde{\Omega}$ inside and outside of $m$. For the charge difference of $\tilde{m}$, we have the possibilities $m_m0$ and $m_mC$, where the charge difference is integer and half-integer, respectively.  These cases correspond to statistics angles $\Phi_{\Omega,m;m}=0,\pi$ ($m_m 0$) and $\Phi_{\Omega,m;m}=\pm \pi/2$ ($m_m C$) in the three loop braiding process shown in Fig.\ref{DRm} (c).  The difference between $0$ and $\pi$, and between $+ \pi/2$ and $-\pi/2$, corresponds to redefining $m_m$ by binding a symmetry charge $Q$.

For the difference in topological spin, the possibilities are $\Omega_m b$ and $\Omega_m s$, where the exchange statistics of $\tilde{\Omega}$ modulo $\pi$ is the same or different (by $\pi/2$) on the two sides of $m$.   These cases correspond to $\Phi_{\Omega,\Omega;m} =0$ ($\Omega_m b$) and $\Phi_{\Omega,\Omega;m} =\pi$ ($\Omega_m s$).

It seems that there are four possibilities in the $e0$ case, but in fact  $e0 m_mC \Omega_m b$  is the same as  $e0 m_mC \Omega_m s$. Starting from one of these cases, if we redefine $\Omega$ by attaching a gauge flux $m$, we change between $\Omega_m b$ and $\Omega_m s$. Due to the $\pm i$ braiding statistics between $m_m$ and $\Omega_m$, such a redefinition changes the braiding statistics between two $\Omega_m$ loops, and hence alters the difference in topological spin of $\tilde{\Omega}$ between the two sides of $m$. This redefinition of $\Omega$ does not change the braiding between $\Omega_m$ and $m_m$, so the charge difference $m_m C$ is not affected.

Of course, such a redefinition applies not only to $\Omega$ loops linked with a base loop $m$, but to all $\Omega$ loops.  We cannot attach quasi-strings differently to different $\Omega$ loops.  This is because any two $\Omega$ and $m \Omega$ loops can be distinguished by braiding an $e$ quasi-particle around each of loop; the statistical phase acquired differs by a factor of $-1$ coming from braiding $e$ around $m$.  This holds regardless of whether the $\Omega$ and $m \Omega$ loops are linked with any other loops.  While it may seem obvious that the redefinition of $\Omega$ by attaching quasi-strings should be independent of base loop, we emphasize it here in order to contrast it to the case of attaching quasi-particles discussed below, which can be different for different base loops.

Therefore, in the $e0m_mC$ case, the topological spin difference of $\t{\Omega}$ is actually not well defined. 
This provides a particular example where the difference in $Z_2^s$ SPT order on the two sides of a $m$ base loop is not well defined due to the nontrivial SET order present. On the other hand, this difference is well defined for the $e0m_m0$ case, and is responsible for the distinction between the $e0m_m0\Omega b$ and $e0 m_m 0 \Omega s$ SF patterns.


Next we consider the case of $eC$, where $e$ carries half integer $Z_2^s$ charge. In this case, $e$ also carries half integer $Z_2^s$ charge in the dimensionally reduced $2D$ SET phase, on both sides of the $m$ base loop. To specify the SF pattern on $m$, we again need to consider the possibilities $m_m0$ / $m_mC$ and $\Omega_m b$ / $\Omega_m s$. 

First we notice that $\Omega_m b$ and $\Omega_m s$ are always equivalent to each other in this case, because we can redefine $\Omega_m$ by attaching an $e$.  This shifts the difference in topological spin of $\tilde{\Omega}$ by $\pm i$.  It is important to note that this redefinition is done only for those $\Omega$ loops linked to a $m$ base loop.  In particular, we do not redefine $\Omega$ loops which are not linked with any base loop. This is consistent, because $\Omega$ loops linked with a $m$ loop cannot be shrunk down to a point, and therefore the quasi-particle type of such loops is not well-defined. One cannot meaningfully compare the quasi-particle type of $\Omega$ loops linked with an $m$ base loop with that of the $\Omega$ loops linked with other base loops. This point was mentioned in \onlinecite{Wang2014a} and discussed in more detail in \onlinecite{Lin2015}.

Hence in the case of $eC$, we only have the distinction between $m_m0$ and $m_mC$. One might try to change between these two possibilities by attaching an $e$ to the $m_m$ loop. However, this redefinition causes $\tilde{m}$ to have different exchange statistics on the two sides of the $m$ base loop.  We already showed in Sec.~\ref{mSET} that $\tilde{m}$ can be taken to have the same statistics everywhere.  Making this choice, the charge difference of $\tilde{m}$ across the $m$ base loop is well-defined, and $eC m_m0$ and $eC m_mC$ describe different SF patterns.

In table \ref{tb_SF}, we list all five possible SF patterns in the $3D$ $Z_2Z_2$ SET.
\begin{table}[htbp]
\begin{tabular}{| c | c | c | c |}
  \hline                       
  SF patterns & $Z_2^s$ charge on $e$ & $Z_2^s$ charge on $m_m$ & topo spin of $\Omega_m$\\
  \hline
  \hline
  $e0m_m0\Omega_m b$ & integer & integer & $\pm 1$\\
  \hline
  $e0m_m0\Omega_m s$ & integer & integer & $\pm i$ \\
  \hline  
  $e0m_mC$ & integer & half integer & $\pm 1$ or $\pm i$ \\
  \hline
  $eCm_m0$ & half integer & integer & $\pm 1$ or $\pm i$\\
  \hline
  $eCm_mC$ & half integer & half integer & $\pm 1$ / $\pm i$\\
  \hline
\end{tabular}
\caption{Symmetry fractionalization patterns for $3D$ $Z_2^g$ gauge theory with $Z_2^s$ symmetry. $e0$ and $eC$ refer to the gauge charge $e$ carrying integer and half integer symmetry charge respectively. $m_m0$ and $m_mC$ refer to the gauge flux loop $m$ carrying integer or half integer symmetry charge when it is linked with a base loop of $m$. They correspond to a phase factor of $\pm 1$ and $\pm i$ in the three loop braiding process shown in Fig.\ref{DRm} (c). $\Omega_m b$ and $\Omega_m s$ refer to the symmetry flux loop having bosonic or semionic topological spin when linked with a base loop of $m$. They correspond to a phase factor of $1$ and $-1$ in the three loop braiding process shown in Fig.\ref{mZ2SPT} (c). In the last three cases, $\Omega_m$ can be either bosonic, fermionic or semionic depending on the quasi-particle or quasi-string attachment.}
\label{tb_SF}
\end{table}

\section{Anomaly detection}
\label{anom}

Can all the SF patterns listed in Table~\ref{tb_SF} be realized? In $2D$ SET phases, we know that some SF patterns may be anomalous, in the sense that they cannot be realized in strictly $2D$ systems, but can be realized at the boundary of a $3D$ SPT phase.  Is this the case for any of the $Z_2Z_2$ SET phases we listed in Table \ref{tb_SF}?

To answer this question, we need an anomaly detection method. Several such methods have been proposed, but they mostly work in $2D$, and generalization to $3D$ is not straight-forward. In a recent paper, we introduced the flux fusion idea for anomaly detection in some $2D$ SET phases,\cite{Hermele2015} which other groups have extended and applied to a wider range of $2D$ SET phases.\cite{Qi2015,Zaletel2015,Cincio2015}  Here, we generalize this idea to $3D$ and and apply it to the $Z_2Z_2$ SET example. In section \ref{FF}, we review the flux fusion idea and use a $2D$ anomalous SET phase as an example to illustrate how it works. In section \ref{FF3}, we generalize the method to $3D$ and show that the $eCm_mC$ $Z_2Z_2$ SET is anomalous. The $eCm_mC$ SET can still be realized, but only as the surface of a $4D$ SPT phase, which we show using a coupled layer construction in section \ref{CL}. In section \ref{NA}, we show that the other four $Z_2Z_2$ SET phases are non-anomalous, by showing that the $Z_2^s$ symmetry can be consistently gauged, resulting in $3D$ gauge theories with larger gauge group.

\subsection{The flux fusion idea: recap}
\label{FF}

The flux fusion method for detecting anomalous SF patterns works as follows:\cite{Hermele2015}
\begin{enumerate}
\item{Use the SF pattern to deduce the fusion rules of symmetry fluxes.}
\item{Consider how symmetry fractionalizes on symmetry fluxes.}
\item{Determine if \emph{any} SF pattern on symmetry fluxes is consistent with the SF pattern of the fractional excitations and the fusion rules of the symmetry fluxes.}
\end{enumerate}
If no consistent SF pattern on symmetry fluxes exists, then an anomaly has been detected.

To illustrate how this works in detail, we consider the example of a $2D$ $Z_2$ gauge theory with $U(1) \times Z_2^T$ symmetry, where $Z_2^T$ is anti-unitary time reversal symmetry.  Note that the $U(1)$ symmetry commutes with  time reversal. Therefore, the $U(1)$ charge is reversed under time reversal, and we can think of the $U(1)$ as  spin rotation. We consider the SF pattern $eCmT$, which means that the gauge charge $e$ carries half $U(1)$ charge and transforms as $T^2 = 1$ under time reversal, while $m$ carries integer $U(1)$ charge and is a Kramers doublet under time reversal ($T^2 = -1$).  We note that $m$ is a quasi-particle here, because this is a $2D$ example.  This SF pattern was argued to be anomalous in~\onlinecite{Wang2013}.  The same conclusion was obtained via the flux fusion method in~\onlinecite{Hermele2015}; we follow this discussion here, and more details can be found in~\onlinecite{Hermele2015}.

\begin{figure}[htbp]
\begin{center}
\includegraphics[width=8.0cm]{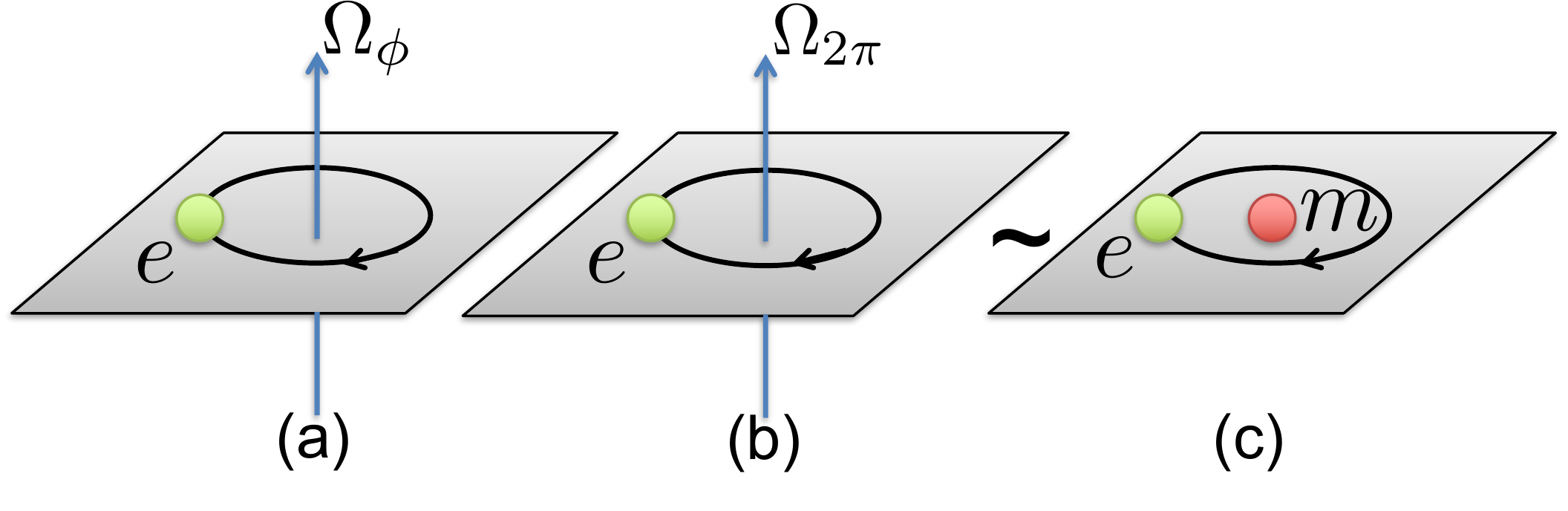}
\caption{Relation between flux fusion and fractional excitations in the $2D$ $eCmT$ theory: (a) braiding an $e$ quasi-particle around a $\phi$ $U(1)$ flux results in a $e^{i\phi/2}$ phase factor; (b) braiding $e$ around a $2\pi$ flux results in a phase factor of $-1$ which is equivalent to (c) braiding $e$ around $m$.}
\label{eCmT}
\end{center}
\end{figure}

In the first step, we insert $U(1)$ flux into the system.  We consider 
 the configuration shown in Fig.\ref{eCmT} (a), where we insert a $\phi$ flux $\Omega_{\phi}$ into the system, and bring an $e$ around it. Because $e$ carries half $U(1)$ charge, a statistical phase of $e^{i \phi/2}$ is accumulated in this braiding process.

Now consider the case of $\phi=2\pi$. A $2\pi$ flux should be equivalent to zero flux; however, bringing an $e$ quasi-particle around it results in a phase factor of $e^{i2\pi/2}=-1$ instead of $1$. Therefore, something nontrivial happens with the $2\pi$ flux. In fact, a $2\pi$ flux in this SET theory is equivalent to an $m$ particle, which explains the $-1$ phase factor when $e$ goes around it. In this way, we complete the first step of the flux fusion method, finding that
\be
\Omega_{2\pi} \sim m \text{.}
\ee

In the second step, we ask about the symmetry fractionalization pattern of the symmetry fluxes. In particular, we ask how  time reversal symmetry $T$ fractionalizes on the $U(1)$ fluxes  $\Omega_\phi$. We note that $T$ leaves the flux unchanged, \emph{i.e.} $T : \Omega_{\phi} \mapsto \Omega_{\phi}$, so it is well-defined to talk about the action of $T^2$ on $\Omega_{\phi}$.  When $\phi=0$, naturally time reversal cannot fractionalize and acts in the usual way with $T^2=1$. Due to the continuity of $\phi$, we would expect this to be true for all $\phi$. Therefore, we find
\be
T^2=1 \text{\ on all $U(1)$ fluxes $\Omega_\phi$}
\ee

In the third step, we realize that this is not consistent with how time reversal fractionalize on the quasi-particles. In particular, time reversal acts as $T^2=-1$ on $m$. However, we found that $\Omega_{2\pi} \sim m$ and $T^2=1$ on $\Omega_{2\pi}$. Hence we have a contradiction, and the $eCmT$ SF pattern is anomalous.

\subsection{Identify $eCm_mC$ as anomalous 3D $Z_2Z_2$ SET}
\label{FF3}

Now we follow the same logic and identify the $eCm_mC$ SF pattern listed in Table \ref{tb_SF} as anomalous. 

As explained in the previous section, in specifying the $m_m C$ SF pattern, we are focusing on the bosonic $m_m$ loop. When an $m$ loop is linked with a base loop of $m$, its topological spin can be either bosonic or fermionic. These two types of $m_m$ loops can be mapped to each other by attaching an $e$ charge. $m_mC$ is the SF pattern where the bosonic $m_m$ loop carries half $Z_2^s$ symmetry charge.

\begin{figure}[htbp]
\begin{center}
\includegraphics[width=8.0cm]{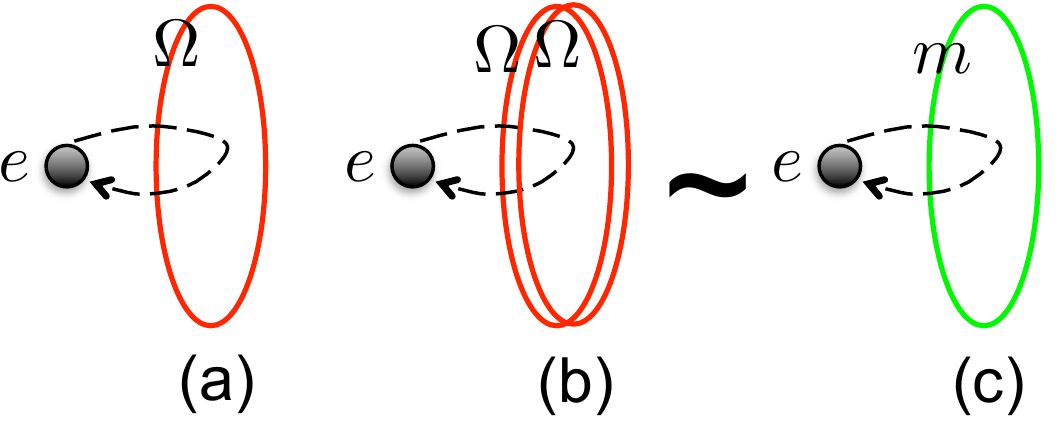}
\caption{Relation between flux fusion and fractional excitations in the $3D$ $eCm_mC$ theory: (a) braiding an $e$ quasi-particle around an $\Omega$ $Z_2^s$ flux loop results in a $\pm i$ phase factor; (b) braiding $e$ around two $\Omega$ flux loops results in a phase factor of $-1$ which is equivalent to (c) braiding $e$ around an $m$ loop.}
\label{OOm}
\end{center}
\end{figure}

First, from the fact that $e$ carries half $Z_2^s$ charge we find that two $Z_2^s$ symmetry fluxes ($\Omega$) fuse into an $m$ loop. To see this, we notice that braiding an $e$ around one $\Omega$ loop results in a phase factor of $\pm i$ and braiding $e$ around two $\Omega$ loops results in a phase factor of $-1$, as shown in Fig.\ref{OOm}. Therefore, the fusion result of two $\Omega$ loops is not vacuum. Instead, it is the $m$ loop. 
\be
\Omega \times \Omega \sim m
\ee

\begin{figure}[htbp]
\begin{center}
\includegraphics[width=8.0cm]{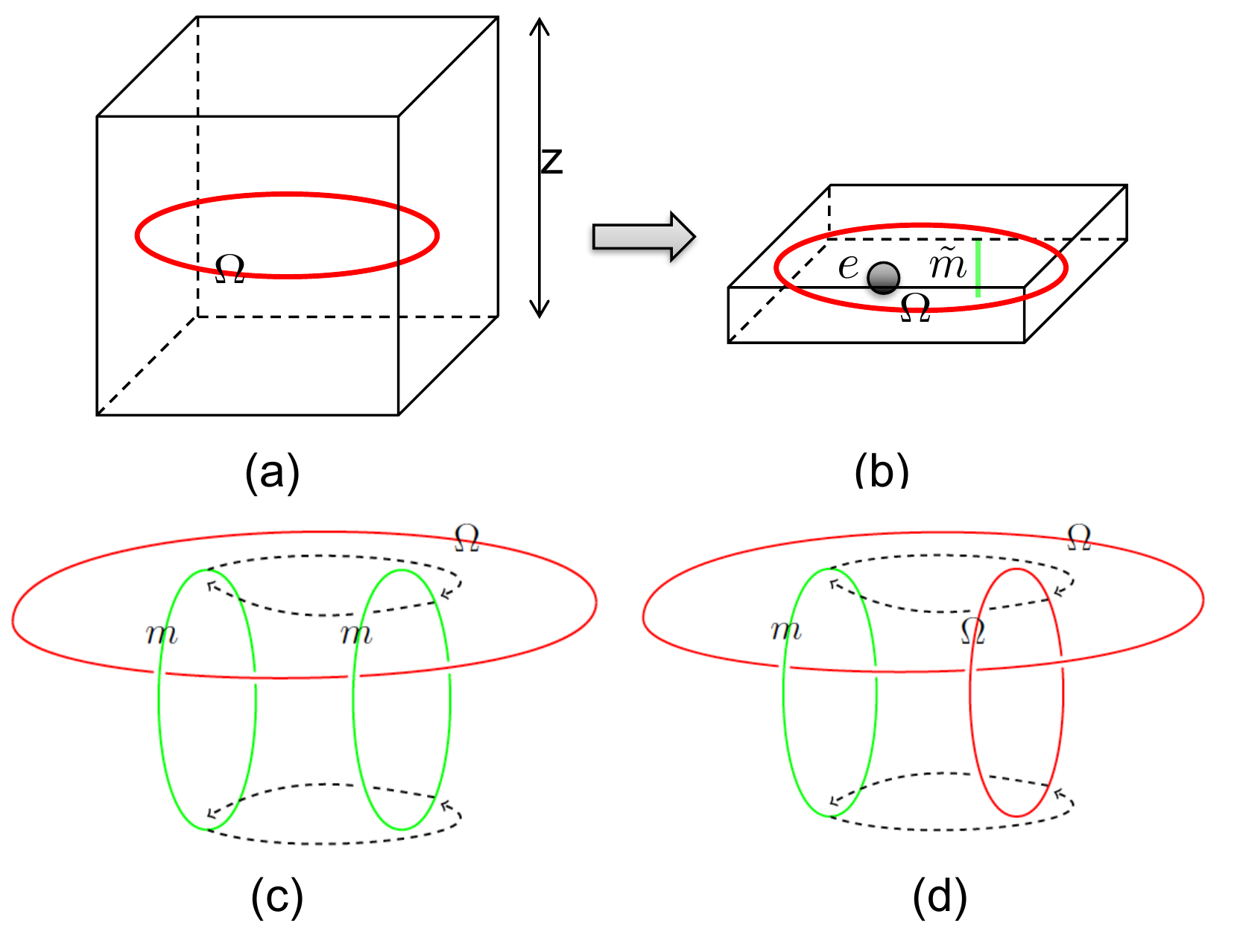}
\caption{Dimensional reduction with an $\Omega$ loop in the $xy$ plane: (a) insert an $\Omega$ loop in the $xy$ plane and (b) compress the system down in the $z$ direction. The $\tilde{m}$ quasi-particle can have different topological spin and different $Z_2^s$ symmetry charge on the inside and outside of $\Omega$ corresponding to nontrivial three loop braiding processes shown in (c) and (d) respectively.}
\label{DRO}
\end{center}
\end{figure}

Next, we ask how  $Z_2^s$ symmetry fractionalizes on the $\Omega$ loop. In order to answer this question, we insert an $\Omega$ loop in the $xy$ plane and compress the system down in the $z$ direction, as shown in Fig.\ref{DRO}. This is similar to the dimensional reduction procedure illustrated in Fig.\ref{DRm}. The only difference is that we are now inserting an $\Omega$ loop instead of an $m$ loop. The $\Omega$ loop can be inserted while preserving $Z_2^s$ symmetry, so after dimensional reduction  we get a $2D$ SET state, with possibly different SET orders inside and outside of the $\Omega$ loop. The difference in SET orders describes the SF pattern on the $\Omega$ loop. 

Compared to the case of dimensional reduction with an $m$ loop,  dimensional reduction with an $\Omega$ loop gives rise to new possibilities.  In particular, we should consider the possibility that the topological orders inside and outside the $\Omega$ loop are different.  The quasi-particles on the two sides of the $\Omega$ base loop are given by $e$, $\tilde{m}$ and their composites under fusion, as shown in Fig.\ref{DRO} (b).  In any region, the topological spin of $\tilde{m}$ can be $\pm 1$ or $\pm i$; this is so because $\tilde{m}^2$ is trivial or a quasi-particle of the $3D$ theory, and in either case is a boson.  The three loop braiding process shown in Fig.\ref{DRO} (c) measures the difference in topological spin of $\tilde{m}$ between the inside and outside of the $\Omega$ base loop, that is
\be
\Phi_{m,m;\Omega} =  \theta^{[\Omega]}_{m,m} - \theta^{[I]}_{m,m} = 0, \pi \text{.}
\ee
We label the two possibilities as $m_{\Omega}b$ and $m_{\Omega} s$, respectively.

Moreover, we should also consider the difference in  $Z_2^s$ symmetry charge  carried by $\tilde{m}$ in the two regions. There are two options: the charges can either be the same (up to an integer charge difference) or differ by a half integer charge. The three loop braiding process shown in Fig.\ref{DRO} (d) results in a phase factor of $\pm 1$ and $\pm i$ respectively in these two cases,
\be
\Phi_{\Omega,m;\Omega}=0, \pi \text{\ or \ } \pm\pi/2 \text{.}
\ee
We label these two possibilities as $m_{\Omega}0$ and $m_{\Omega}C$. 

For the third step, we want to see if any of the SF patterns on $\Omega$ can be consistent with the SF pattern of $m$, given the fusion rule $\Omega \times \Omega \sim m$. It turns out that none of the SF patterns are consistent, and this is how we detect an anomaly.

First, we show that $m_{\Omega}s$ is not possible. Because two $\Omega$ loops fuse into an $m$ loop, upon gauging the symmetry we obtain a $Z_4$ gauge theory in $3D$.  It is shown in Ref.~\onlinecite{Wang2014a} that $3 \cdot 4 \Phi_{\Omega;\Omega} = 0 \operatorname{mod} 2\pi$, so that $\Phi_{\Omega;\Omega} = 2 \pi k / 12$ for integer $k$.  Consider two $\Omega$ loops, both linked to the same $\Omega$ base loop, with the same topological spin (\emph{i.e.} both characterized by the same value of $k$).  Fusing these two loops together gives a $m_{\Omega}$ loop with exchange statistics $\Phi_{m;\Omega} = 2 \pi k / 3$.  Fusing two such $m_{\Omega}$ loops gives a quasi-particle excitation, with statistics $8 \pi k / 3$.  But all quasi-particles are bosons, implying that $k = 0 \operatorname{mod} 3$.  Therefore, $\Phi_{\Omega; \Omega} = 2 \pi k' / 4$, for integer $k'$, the topological spin of $m_{\Omega}$ is always $1$.   If we take into account of the possibilities of attaching quasi-particles to $m_{\Omega}$, its topological spin can at most be $-1$. Therefore, $m_{\Omega}s$ is not consistent with $\Omega \times \Omega \sim m$ and only $m_{\Omega}b$ is possible.

Now we consider the possibilities of $m_{\Omega}0$ and $m_{\Omega}C$. To clarify the meaning of these two possibilities, we need to specify which $m_{\Omega}$ loop we are looking at, because by attaching an $e$ charge we can change the fractional symmetry charge carried by the loop. Here we use the convention that $m_{\Omega} 0$ and $m_{\Omega} C$ refers to the fractional symmetry charge carried by the bosonic $m_{\Omega}$ loop. If we attach an $e$ charge to it, the fractional charge carried by the loop changes, and at the same time the topological spin of the loop changes to fermionic, due to the $-1$ phase factor resulting from the braiding of $e$ around $m_{\Omega}$.

Now we show that neither  $m_{\Omega}0$ nor $m_{\Omega}C$  is consistent with the $eCm_mC$ SF pattern on the $m$ loop. We observe that, independent of the $Z_2^s$ charge of $m_{\Omega}$, $m_m$ always carries integer $Z_2^s$ charge.  This follows from the fact that the $m$ base loop can be decomposed into two $\Omega$ base loops, and the linearity of three-loop braiding statistics.  In particular we have
\be
\Phi_{m,\Omega;m} = \Phi_{m,\Omega;\Omega} + \Phi_{m,\Omega;\Omega} = 2 \Phi_{m,\Omega;\Omega} \text{.}
\ee
Because $\Phi_{m,\Omega;\Omega} = q \pi /2$ for $q = 0,\dots,3$, this implies $\Phi_{m,\Omega;m} = \pi q$.
Moreover, as we have chosen the $m$ loops linked with $\Omega$ to be bosonic, their composite $m$ loop linked with $\Omega \times \Omega= m$ is also bosonic. We have thus obtained a contradiction with the $m_mC$ SF pattern, which says that the bosonic $m_m$ loop carries half-integer $Z_2^s$ charge. Therefore, none of the possible SF patterns on $\Omega$ is consistent with the SF pattern on $m$ and we have found an anomaly.

\subsection{Other $Z_2Z_2$ SETs are non-anomalous}
\label{NA}

While the $eCm_mC$ $Z_2Z_2$ SET phase is anomalous, the other four listed in Table \ref{tb_SF} are not. In fact, the $Z_2^s$ symmetry in these SET phases can be consistently gauged, resulting in a larger topological theory in $3D$. We discuss each of these four cases in the following.

First, the $e0m_m0\Omega_mb$ case is the simplest, with only trivial $Z_2^s$ symmetry action on the fractional excitations ($e$ or $m$). Therefore, the $Z_2^g$ gauge theory  and  $Z_2^s$ symmetry sectors are independent of each other. After gauging, we obtain a $Z_2  \times Z_2$ gauge theory which is just composed of two copies of the usual $Z_2$ gauge theory.

Next, we consider the $eCm_m0$ case. In this case, $e$ transforms nontrivially under the $Z_2^s$ symmetry, but $m$ is trivial. Using the argument presented in section \ref{FF3}, we find that two $Z_2^s$ symmetry fluxes $\Omega$ fuse into one $m$ loop. Therefore, if the $Z_2^s$ symmetry is gauged, we obtain a $Z_4$ gauge theory with $\Omega$ being the fundamental flux loop. $e$  becomes the fundamental gauge charge. The $Z_4$ gauge theory has only trivial loop braiding processes.

Now let's consider the case of $e0m_m0\Omega_m s$. As $e$ carries integer charge under $Z_2^s$, two $\Omega$ loops should fuse into vacuum instead of the $m$ loop, therefore, we would expect to get some $Z_2\times Z_2$ gauge theory. However, this is a twisted $Z_2\times Z_2$ gauge theory, due to the nontrivial three loop braiding process between two $\Omega$ loops with base loop $m$. This braiding process results in a phase factor of $-1$; that is,
\be
\Phi_{\Omega,\Omega;m} = \pi \text{,}
\ee
which is equivalent to saying that the topological spin of $\Omega$ when linked with $m$ is $\pm i$.  The other three loop braiding processes with base loop $m$ are trivial. Given this information, the general constraints on three loop braiding obtained in Ref.~\onlinecite{Wang2014a} imply that braiding of two $\Omega$ loops linked with an $\Omega$ base loop is trivial, and that
\be
\Phi_{\Omega,m;\Omega} = \pm \pi/2
\ee
The same three loop braiding statistics were shown in~\onlinecite{Wang2014a,Mesaros2013} to be realized in a model for a bosonic SPT phase with $Z_2 \times Z_2$ global symmetry, upon gauging the full symmetry.  This implies that we can realize the $e0m_m0\Omega_m s$ SET phase by starting with such a bosonic SPT phase, and gauging one of the $Z_2$ factors, leaving the remaining factor as the $Z_2^s$ global symmetry.  This provides a route to realize this SET phase via parton constructions, by putting bosonic partons into an appropriate SPT phase, and may be helpful in identifying physically reasonable realizing this non-trivial SET phase. 

Finally, we consider $e0m_mC$. Similar to the previous case, we expect to get a $Z_2\times Z_2$ gauge theory, but also with some twisting. In particular, due to the fractional $Z_2^s$ charge carried by $m_m$ when linked with base loop $m$, there is non-trivial braiding between $m$ and $\Omega$ when linked with $m$:
\be
\Phi_{m,\Omega;m}=\pm \pi/2 \text{.}
\ee
In addition, we argued in Sec.~\ref{sym_frac} that we can take $\Omega_m$ to be bosonic when linked with $m$, so that 
\begin{equation}
\Phi_{\Omega,\Omega;m} = 0 \text{.}
\end{equation}
This information, combined with the general constraints of~\onlinecite{Wang2014a}, implies
\be
\Phi_{m,m;\Omega}= \pi \text{.}
\ee
and that braiding of two $\Omega$ loops linked with a $\Omega$ base loop is trivial.  Again, the same three loop braiding was obtained in Ref.~\onlinecite{Wang2014a,Mesaros2013} by gauging a SPT phase with $Z_2 \times Z_2$ symmetry.

In Table \ref{tb_g}, we list all the non-anomalous $Z_2Z_2$ SET phases, and the corresponding gauge theory when the $Z_2^s$ symmetry is gauged. \footnote{Note that there are four $Z_2\times Z_2$ gauge theories in $3D$ (including twisted and untwisted) while in table \ref{tb_g} only three of them are listed. In fact, the fourth one can be obtained from the one in the third row by redefining $\Omega$ as $m\Omega$.}

\begin{table}[htbp]
\begin{tabular}{ | c | c | c | }
  \hline                       
  $Z_2Z_2$ SET & Gauging result & Nontrivial 3 loop braiding\\
  \hline
  \hline
  $e0m_m0\Omega_m b$ & untwisted $Z_2\times Z_2$ GT & none \\
  \hline
  $e0m_m0\Omega_m s$ & twisted $Z_2\times Z_2$ GT & $\Phi_{\Omega,\Omega;m} = -1$ \\
  \hline 
  $e0m_mC$ & twisted $Z_2\times Z_2$ GT & $\Phi_{m,\Omega;m}=\pm i$ \\
  \hline
  $eCm_m0$ & $Z_4$ gauge theory & none \\
  \hline
\end{tabular}
\caption{Non-anomalous $Z_2Z_2$ SETs and their corresponding gauge theory when the $Z_2^s$ symmetry is gauged. Nontrivial three loop braiding statistics in the gauge theory is also listed. GT stands for gauge theory.}
\label{tb_g}
\end{table}

\subsection{Anomalous $Z_2Z_2$ SET as surface of $4D$ $Z_2$ SPT}
\label{CL}

The $eCm_mC$ $Z_2Z_2$ SET is anomalous and cannot be realized in a strictly $3D$ system. However, it can be realized on the surface of a $4D$ system. This is similar to  anomalous SETs  studied in $2D$, which can be realized as the surface of a $3D$ system. In this section we present a `coupled layer' construction of such a $4D$ system realizing the $eCm_m C$ SF pattern on its surface, following a similar construction introduced to realized $2D$ anomalous SETs.\cite{Wang2013} Here each ``layer'' used in the construction is actually a $3D$ system.

\begin{figure}[htbp]
\begin{center}
\includegraphics[width=5.0cm]{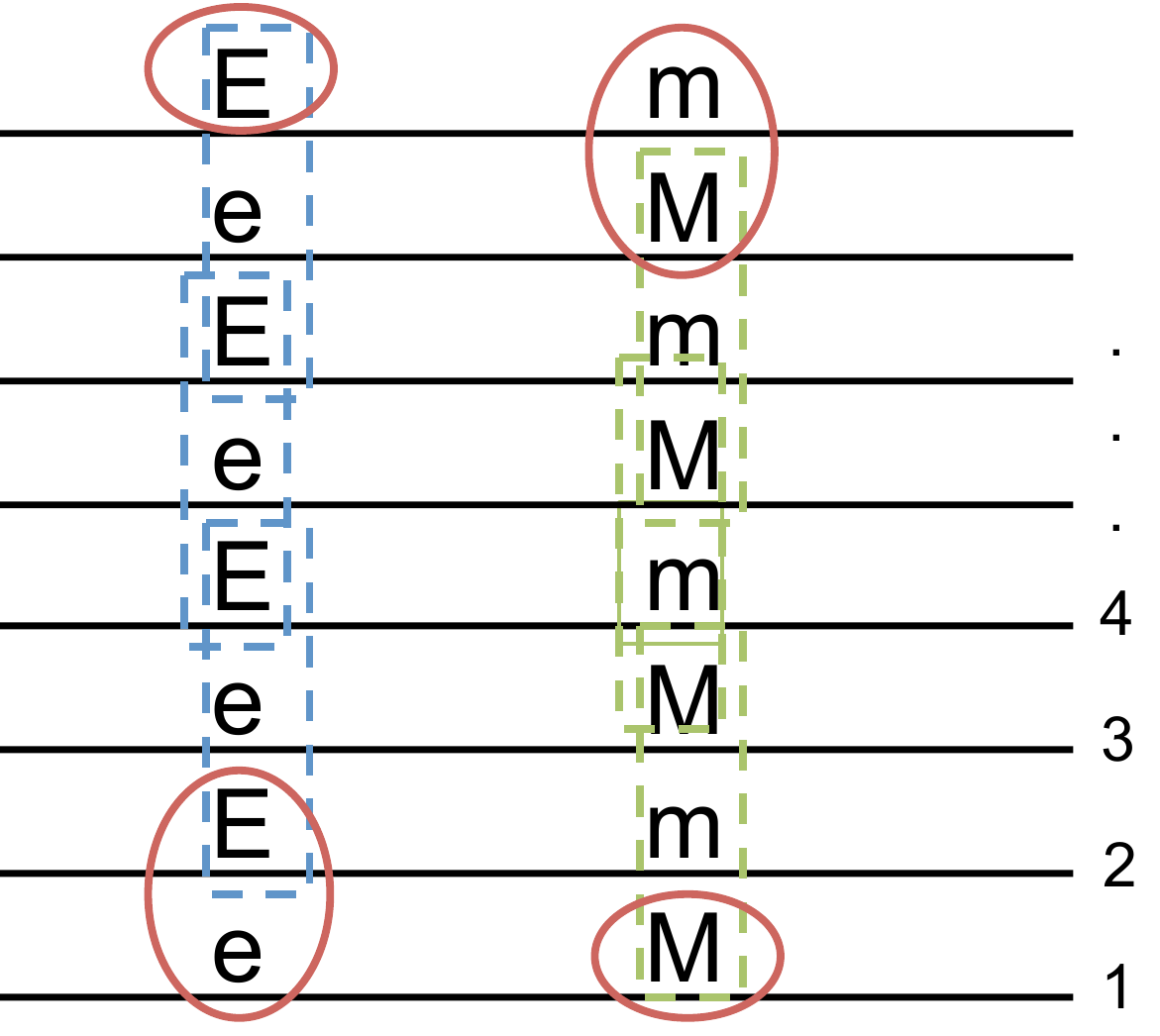}
\caption{Coupled layer construction of a $4D$ system realizing the $eCm_mC$ anomalous SET on its surface. Each layer is a $3D$ $Z_2^g$ gauge theory. $e$ and $E$ are $Z_2^g$ gauge charges without and with fractional $Z_2^s$ charge respectively. $m$ and $M$ are $Z_2^g$ gauge fluxes without and with nontrivial $Z_2^s$ symmetry fractionalization. Fractional excitations in the dotted boxes are condensed and those in red circles remain after the condensation as surface anomalous $Z_2Z_2$ SET.}
\label{CL4}
\end{center}
\end{figure}

As shown in Fig.~\ref{CL4}, each layer is a $3D$ $Z_2^g$ gauge theory with $Z_2^s$ symmetry. $e$ and $E$ denote $Z_2^g$ gauge charges and $m$ and $M$ denote $Z_2^g$ gauge fluxes. $e$ transforms trivially under $Z_2^s$ while $E$ carries fractional charge of $Z_2^s$. Also $m$ transforms trivially under $Z_2^s$ while $M$ transforms in the same way as the flux loop in $eCm_mC$. That is, the $Z_2Z_2$ SET in the odd-numbered layers (counting from the bottom) has the $e0m_mC$  SF pattern and those in the even-numbered layers have the $eCm_m0\Omega_m b$ SF pattern. Therefore, each layer is made up of non-anomalous $Z_2Z_2$ SETs which can be realized in strictly three dimensions.

Now we condense the composite objects in the dashed boxes as shown in Fig.~\ref{CL4}. Note that the composite inside each box is either a bosonic quasi-particle or a quasi-string with trivial exchange statistics, whether linked with a base loop or not. Therefore, they can be condensed. Moreover, this can be done without breaking the $Z_2^s$ symmetry. This is because each box always contains either two $E$'s or two $M$'s. Therefore, even though each $E$ or $M$ transforms nontrivially under $Z_2^s$, the composite inside each box always transforms trivially. After such condensation, we can check explicitly that all the fractional excitations in the bulk (both quasi-particles and quasi-strings) are either condensed or confined. On the surface, there are non-trivial excitations left behind, which are indicated by red circles in Fig.~\ref{CL4}. On the top surface, the remaining fractional excitations are $E$ and the composite $mM$ on neighboring layers. These are bosonic quasi-particle and bosonic quasi-string excitations and have 
mutual $-1$ braiding statistics. Therefore, the top surface has a $3D$ $Z_2^g$ topological order. The quasi-particle $E$ carries half $Z_2^s$ charge. Moreover, when a $mM$ loop is linked with another $mM$ loop, it has half $Z_2^s$ charge. Therefore, we realize the $eCm_mC$ SF pattern on the top surface. Similar arguments show that the same SF pattern is realized on the bottom surface.

Therefore, the $eCm_mC$ $Z_2Z_2$ SET can be realized as the surface state of a $4D$ system. The bulk of the $4D$ system should have a nontrivial $Z_2^s$ symmetry protected topological order to cancel the anomaly coming from the surface.

\section{Summary and Discussion}
\label{sum}

In this paper, we studied three dimensional topological phases with symmetry and addressed the following questions:
\begin{enumerate}
\item{How to describe symmetry fractionalization patterns in $3D$?}
\item{How to detect anomalies in the symmetry fractionalization patterns?}
\end{enumerate}

In answering the first question, we found that
\begin{enumerate}
\item{The SF pattern of a quasi-particle is given by a fractional representation of the symmetry, similar to the $2D$ case.}
\item{The SF pattern of a quasi-string is given by the difference of two $2D$ SET orders, one on each side of the quasi-string in the dimensionally reduced $2D$ plane containing the quasi-string.}
\end{enumerate}

Moreover, in combining the SF patterns of quasi-particles and quasi-strings into a full description of the SF pattern in $3D$ topological phases, we need to satisfy the following conditions:
\begin{enumerate}
\item{Consistency: The SF pattern of quasi-particles should be the same both in the $3D$ bulk and in the dimensionally reduced systems used for the description of  the SF pattern on quasi-strings.}
\item{Redundancy: Two SF patterns are the same if they
can be related by two kinds of operations:
(a) Redefining symmetry fluxes
when they are linked with nontrivial base loops
by attaching quasi-particles to them. Such quasiparticle
attachment can be different for different
base loops. (b) Redefining symmetry fluxes by attaching
quasi-strings to them. Such quasi-string
attachment should be independent of the base loop.}
\end{enumerate}

In particular, we find that for $3D$ $Z_2^g$ gauge theory with $Z_2^s$ symmetry, there are five different SF patterns, as listed in Table \ref{tb_SF}. 

In answering the second question, we employ the flux fusion method introduced in \onlinecite{Hermele2015}. The steps are summarized as follows:
\begin{enumerate}
\item{Use the SF pattern to deduce the fusion rules of symmetry fluxes.}
\item{Consider how symmetry fractionalizes on symmetry fluxes.}
\item{Determine if \emph{any} SF pattern on symmetry fluxes is consistent with the SF pattern of the fractional excitations and the fusion rules of the symmetry fluxes.}
\end{enumerate}
An anomaly in SF pattern is detected if no consistent SF pattern on symmetry fluxes exists.

The flux fusion idea applies only to a limited set of SET orders, but it does work in our case of $Z_2^g$ gauge theory with $Z_2^s$ symmetry, and allows us to identify an anomaly in one of the five SF patterns identified above. While this pattern cannot be realized strictly in $3D$, it is shown to be realizable as the surface of a $4D$ system. We find that the other four SF patterns are non-anomalous, and the $Z_2^s$ symmetry can be consistently gauged.

Many questions are still open regarding $3D$ symmetry enriched topological phases.

First, it is possible that there are SF patterns on quasi-strings that cannot be captured by the dimensional reduction procedure. This is the case for SF patterns on quasi-particles. For example, consider a quasi-particle carrying fractional charge under $Z_2^s$ symmetry. After dimensional reduction as discussed in Fig.\ref{DRe}, we get a $1D$ gapped state with $Z_2^s$ symmetry and the quasi-particle exists as a boundary between two parts of the system. Because there are no nontrivial SPT orders in $1D$ with $Z_2^s$ symmetry, the whole $1D$ system is in the same phase, and we do not see any nontrivial features of the quasi-particle. The property of fractional charge carried by the quasi-particle is lost. Of course, in the original bulk of the system, such fractional charge can be  detected by bringing the quasi-particle around a symmetry flux. However, such braiding process is intrinsic to the bulk dimension and is not well-defined upon dimensional reduction to $1D$.

In studying the SF pattern on quasi-strings, we relied on the dimensional reduction procedure. Therefore, similarly, information could have been lost; that is, there could be nontrivial symmetry actions on the quasi-strings that become trivial after dimensional reduction. Such symmetry actions would be related to hypothetical loop braiding processes that are intrinsic to $3D$, in that they do not survive dimensional reduction.  By contrast, the three loop braiding process can be described in terms of dimensional reduction to $2D$ quasi-particle braiding processes on a defect plane.  It is not known if there are loop braiding processes in $3D$ beyond three-loop braiding.  Therefore, our study of $3D$ SET phases is limited by our understanding of $3D$ topological order (loop statistics in particular).

Secondly, it is not clear how to detect anomalies in $3D$ SF patterns in general. The flux fusion method works well for our example, but the method has limitations. For example, it is not known whether it can be usefully extended to handle general nonabelian groups, or nonabelian actions of symmetries on fractional excitations.\cite{Hermele2015} In $2D$, a more general anomaly detection method is known, based on the mathematical framework of G-crossed fusion categories.\cite{Etingof2010} Does a similar framework exist in $3D$? This is a challenging question, given that we do not yet fully understand what topological orders exist in $3D$. However, it might be possible to partially answer this question by restricting attention to $3D$ SET phases with topological orders we already understand, such as untwisted and twisted gauge theories. We leave these questions for future study.

As we were finalizing this manuscript for posting on the arXiv, we noticed a recent preprint\cite{Cheng2015} in which some of the SET phases discussed here were also studied.

\acknowledgments

X.C. would like to acknowledge discussions with Meng Cheng, Ashvin Vishwanath, Michael Levin and Chenjie Wang, Beni Yoshida, Aleksander Kubica. X.C. is supported by  the Caltech Institute for Quantum Information and Matter and the Walter Burke Institute for Theoretical Physics. M.H. is supported by the U.S. Department of Energy, Office of Science, Basic Energy Sciences, under Award number DE-FG02-10ER46686 (April 2015 and earlier) and under Award number DE-SC0014415 (August 2015 and later), and by Simons Foundation grant No.  305008 (sabbatical support).


%

\appendix

\section*{Appendix A: Dimensional reduction of $3D$ $Z_2^g$ gauge theory to $2D$}
\label{app_a}

In this appendix, we examine in more detail the dimension reduction procedure of $3D$ $Z_2^g$ gauge theory to $2D$, justifying the conclusions made in the main text (section \ref{mSET}).  We consider a bosonic system in $3D$, with $Z_2$ topological order (as in the deconfined phase of $\zz$ gauge theory).  The topologically non-trivial excitations are the point particle $e$, which we assume  to be a boson, and the quasi-string $m$.  Here, we do not consider any symmetry.  In Appendix B, we give a similar discussion including $Z_2^s$ symmetry and examining in detail the SET orders upon dimensional reduction to $2D$. 

We start by discussing fusion and braiding properties in $3D$. Besides the three loop braiding parameters $\Phi_{m,m;m}$ and $\Phi_{m;m}$ discussed in the main text, we introduce the following statistical parameters:
\begin{eqnarray}
\begin{array}{ll}
\phi_e = 0 &  \text{self-statistics of } e \\
\phi_{e,m} = \pi & \text{point-loop mutual statistics of } e \text{ and } m \text{.}\\
\end{array}
\end{eqnarray}
Here, $\phi_e$ is the usual statistical angle for the exchange statistics of two $e$ particles.  $\phi_{e,m}$ corresponds to a process where $e$ is brought around a $m$ loop and returned to its original position.  

The three loop braiding parameters are given by
\begin{eqnarray}
\begin{array}{l}
\Phi_{m;m}= 0 \text{\ or\ } \pi\\
\Phi_{m,m;m} = 0\\
\end{array}
\end{eqnarray}
where $\Phi_{m;m}$ denotes the exchange statistics of two $m$ loops linked with a third $m$ base loop and $\Phi_{m,m;m}$ denotes the full braiding between two identical $m$ loops linked with a third base $m$ loop. Note that we can redefine the two linked loops by attaching an $e$ particle to each, $m \to e m$, which shifts $\Phi_{m;m} \to \Phi_{m;m} + \pi$.  Therefore $\Phi_{m;m}$ is only well-defined modulo $\pi$.  In addition, $\Phi_{m,m;m} = 2 \Phi_{m;m} = 0$, so that $\Phi_{m;m} = 0,\pi$,  so we gain no additional information by considering the half-braid three-loop process.

In addition to these braiding processes, we have the fusion rules $e^2 = 1$, and $m^2 = 1$.  The meaning of the $e^2 = 1$ fusion rule is familiar.  The $m^2 = 1$ fusion rule means that if we fuse two $m$ quasi-strings together, we get something that is trivial as a quasi-string.  However, there could  be non-trivial $e$ charge upon fusing two $m$ loops.

Before proceeding, as a brief aside, we would like to establish some consistent notation for different types of braiding processes in $3D$ and $2D$, since below and in the following appendices we will have to consider many different such processes.  For the purposes of the present discussion, we label point particles by $a,b,c,\dots$, and quasi-strings by $\alpha, \beta, \gamma, \dots$. We introduce the following symbols:
\be
\begin{array}{l}
\phi_a:\text{self-statistics of } a \text{ in } 3D \\
\theta_a:\text{self-statistics of } a \text{ in } 2D \\
\phi_{a,\alpha}:\text{point-loop statistics of } a \text{ with } \alpha \text{ in } 3D \\
\theta_{a,b}:\text{mutual statistics of } a \text{ and } b \text{ in } 2D \\
\Phi_{\alpha; \beta}:\text{3-loop braiding (half braid) of two } \alpha \\\text{ loops linked with base loop } \beta \\
\Phi_{\alpha, \beta; \gamma}:\text{3-loop braiding (full braid) of } \alpha \text{ with } \beta \text{,} \\ \text{linked with base loop } \gamma \text{.}
\end{array}
\ee
We use $\theta$ for $2D$ statistics, and $\phi$ and $\Phi$ for $3D$ statistics.  In addition, we use lowercase letters for particle-particle or particle-loop statistics, and capital letters for three loop braiding statistics.

Now we describe dimensional reduction to a $2D$ topologically ordered phase. We always consider periodic boundary conditions in the finite direction. We can reduce either onto vacuum or onto the plane of a $m$ loop, to get two potentially different $2D$ phases.  When we dimensionally reduce onto the plane of $m$ (or of some other quasi-string), we refer to this as the \emph{basal plane}.  We label the $2D$ phases by the quasi-string type of the basal plane (this is $I$ when we dimensionally reduce onto vacuum). The dependence of $2D$ statistics on the basal plane is indicated by a superscript; for example, $\theta_{\tm}^{[m]}$ is the self-statistics of a quasi-particle $\tm$ on the $m$ basal plane.

The dimensionally reduced version of $e$ is again a point particle, which we still refer to as $e$. As discussed in Sec.~\ref{mSET}, dimensional reduction of $m$ quasi-strings is more interesting, because these can stretch across the system in the finite direction, effectively becoming point particles in $2D$.  These point particles are referred to as $\tm$. 

The following fusion and braiding properties hold independent of basal plane:
\begin{eqnarray}
e^2 &=& 1 \\
\theta_e &=& 0 \\
\theta_{e, \tilde{m}} &=& \pi \text{.}
\end{eqnarray}

In principle, we could have either $\tm^2 = 1$ or $\tm^2 = e$.  Two $\tm$'s must fuse to something that is trivial as a quasi-string, but we cannot immediately rule out that they could fuse to $e$, the only non-trivial $2D$ particle which is trivial as a quasi-string.  Note that in either case, $\tm^2$ is a boson, so
\begin{equation}
4 \theta_{\tm} = \theta_{\tm^2} = 0 \text{,}
\end{equation}
which implies 
\begin{equation}
\theta_{\tm} = \frac{\pi}{2} q , \qquad q = 0,\dots,3 \text{.}
\end{equation}
Suppose $\tm^2 = e$, then we have
\begin{equation}
\pi = \theta_{\tm, \tm^2} = 2 \theta_{\tm, \tm} = 4 \theta_{\tm} = 0 \text{,}
\end{equation}
a contradiction.  So independent of basal plane, we must have
\begin{equation}
\tm^2 = 1 \text{.}
\end{equation}

The self-statistics of $\tilde{m}$ may depend not only on the basal plane, but also on the details of the dimensional reduction procedure.  The intuition for this is that the phase obtained upon exchanging two $\tilde{m}$'s can quite naturally depend on the ``height'' of these objects, that is on the system size in the finite direction.  However, we have the following relationship with 3-loop braiding in $3D$,
\begin{equation}
\theta^{[m]}_{\tilde{m}, \tilde{m}} - \theta^{[1]}_{\tilde{m}, \tilde{m}} = \Phi_{m,m;m} = 0 \text{.}
\end{equation}
This implies
\begin{equation}
\theta^{[m]}_{\tilde{m}} - \theta^{[1]}_{\tilde{m}} = 0, \pi \text{.}
\end{equation}
Next, note that $\tilde{m}$ can be redefined by $\tilde{m} \to e \tilde{m}$, and this can be done differently for different basal planes.  This redefinition shifts $\theta_{\tilde{m}} \to \theta_{\tilde{m}} + \pi$, so we can always choose
\begin{equation}
\theta_{\tilde{m}} \equiv \theta^{[m]}_{\tilde{m}} = \theta^{[1]}_{\tilde{m}} \text{.}
\end{equation}
In addition, $\theta_{\tilde{m}}$ itself can be shifted by $\pi$, and up to such shifts the most general choice consistent with $\tilde{m}^2 = 1$ is
\begin{equation}
\theta_{\tilde{m}} = 0, \pi/2 \text{.}
\end{equation}
These choices are distinct, with the former corresponding to the toric code, and the latter to the double semion theory. 

\section*{Appendix B: Dimensional reduction of $3D$ $Z_2^g$ gauge theory with $Z_2^s$ global symmetry to $2D$}
\label{app_b}

Following the approach of Appendix A, we now add $Z_2^s$ on-site unitary symmetry, to consider $Z_2Z_2$ SET phases in $3D$. We give a detailed description of the possible SET orders upon reduction to $2D$. We proceed by gauging $Z_2^s$, which introduces two new objects into our description.  These are the symmetry charge $Q$, which is a point particle, and the symmetry flux $\Omega$, which is a quasi-string.

We first study fusion rules, point-point, and point-loop statistics in the gauged theory.  The fusion rules are 
\begin{eqnarray}
Q^2 &=& 1 \\
m^2 &=& 1 \\
e^2 &=& 1 \text{ or } Q \\
\Omega^2 &=& 1 \text{ or } m \text{.}
\end{eqnarray}
The first two fusion rules are fixed, while the last two depend on the properties of the $Z_2Z_2$ SET phase.  The case $e^2 = 1$ corresponds to ``integer'' $Z_2^s$ charge of $e$, which we denote by $e0$.  Similarly, the case $e^2 = Q$ corresponds to ``half-integer'' $Z_2^s$ charge of $e$, and this is denoted $eC$.

Turning to statistics, the point particles have exchange statistics
\begin{equation}
\phi_e = \phi_Q =  0 \text{.}
\end{equation}
$\phi_e = 0$ by assumption, and $\phi_Q = 0$ because $Q$ is a trivial, local excitation before gauging, so it must have bosonic self-statistics after gauging.

For point-loop statistics, we have
\begin{eqnarray}
\phi_{e,m} &=& \pi \\
\phi_{Q,m} &=& 0 \\
\phi_{Q,\Omega} &=& \pi \text{.}
\end{eqnarray}
$\phi_{Q,m} = 0$ because any other choice would contradict $Q$ being a trivial excitation before gauging.  Moreover, $\phi_{Q,\Omega} = \pi$ is a defining property of the flux line $\Omega$.  There is also the point-loop statistical angle $\phi_{e,\Omega}$, which we now relate to the fusion rules.

First, we consider the $e0$ case, that is assume $e^2 = 1$.  Then
\begin{equation}
0 = \phi_{e^2, \Omega} = 2 \phi_{e, \Omega} \text{,}
\end{equation}
which implies
\begin{equation}
\phi_{e, \Omega} = 0, \pi \text{.}
\end{equation}
If $\phi_{e, \Omega} = \pi$, we can redefine $e \to Q e$, which sets $\phi_{e, \Omega} = 0$ while leaving the other statistics angles and fusion rules unchanged.  Then we have
\begin{equation}
\phi_{e, \Omega^2} = 2 \phi_{e, \Omega} = 0 \text{,}
\end{equation}
which fixes the $\Omega^2 = 1$ fusion rule.

Next, we consider $eC$, that is $e^2 = Q$.  Then
\begin{equation}
\pi = \phi_{e^2, \Omega} = 2 \phi_{e, \Omega} \text{,}
\end{equation}
which implies
\begin{equation}
\phi_{e, \Omega} = \pi/2, 3 \pi / 2 \text{.}
\end{equation}
Again, by redefining $e \to Q e$ if needed, we can choose $\phi_{e,\Omega} = \pi/2$.  Then we have 
\begin{equation}
\phi_{e, \Omega^2} = 2 \phi_{e, \Omega} = \pi \text{,}
\end{equation}
which fixes the $\Omega^2 = m$ fusion rule.

Therefore, choosing $e0$ or $eC$ completely fixes the fusion rules, point-point, and point-loop statistics.  To summarize the two cases, for $e0$ we have
\begin{equation}
\begin{array}{ll}
Q^2 = 1 & \phi_e = \phi_Q = 0 \\
e^2 = 1 & \phi_{e,m} = \phi_{Q,\Omega}  = \pi \\
\Omega^2 = 1 & \phi_{Q,m} = 0 \\
m^2 = 1 & \phi_{e,\Omega} = 0 \text{,}
\end{array}
\end{equation}
while for $eC$:
\begin{equation}
\begin{array}{ll}
Q^2 = 1 & \phi_e = \phi_Q = 0 \\
e^2 = Q & \phi_{e,m} = \phi_{Q,\Omega}  = \pi \\
\Omega^2 = m & \phi_{Q,m} = 0 \\
m^2 = 1 & \phi_{e,\Omega} = \pi/2 \text{.}
\end{array}
\end{equation}

So far we have not said anything about the action of symmetry on $m$, or about three-loop braiding in the gauged theory.  Our goal is to describe the symmetry action on $m$, and we will do this by dimensional reduction to a $2D$ SET phase with $Z_2^s$ symmetry.  Those properties not having to do with symmetry have already been discussed in Appendix A.

To describe the symmetry action on $m$, we will need to consider dimensional reduction onto the vacuum, and onto the $m$ basal plane.  Let us first describe the general properties of the dimensionally reduced theory for some fixed basal plane (we thus do not worry about basal plane labels for the moment).  The particles of the theory are $e, Q, \tm, \tOmega$.  The properties of the latter two particles can depend on the choice of basal plane, so if we want to include basal plane labels, we should write $\tOmega_I$, $\tm_m$, and so on.

Let's describe what we know about the fusion rules and statistics of the dimensionally reduced theory, first for the $e0$ case.  The fusion rules are
\begin{eqnarray}
e^2 &=& 1 \\
Q^2 &=& 1 \\
\tm^2 &=& 1,Q \\
\tOmega^2 &=& 1,e,Q,eQ \text{.}
\end{eqnarray}
The last two fusion rules are partially undetermined so far.  We know  that in $3D$, $m^2 = 1$ and $\Omega^2 = 1$, which means, for example, that two $\tm$'s cannot fuse to the dimensional reduction of a non-trivial quasi-string.  We also already showed in Appendix A that that two $\tm$'s cannot fuse to $e$ (or to $eQ$ in the gauged theory).  The statistics are
\begin{eqnarray}
\theta_e &=& \theta_Q = \theta_{e,Q} = \theta_{Q, \tm} = 0 \\
\theta_{e,\tm} &=& \theta_{Q, \tOmega} = \pi \\
\theta_{e,\tOmega} &=& 0 \\
\theta_{\tm} &=& 0, \pi/2 = \text{?} \\
\theta_{\tOmega} &=&  \text{?} \\
\theta_{\tm,\tOmega} &=& \text{?} \text{.}
\end{eqnarray}
We recall from Appendix A that $\theta_{\tm} = \theta_{\tm_I} = \theta_{\tm_m} = 0,\pi/2$, while the other unknown parameters may depend on basal plane.

In the $eC$ case we have for the fusion rules
\begin{eqnarray}
e^2 &=& Q \\
Q^2 &=& 1 \\
\tm^2 &=& 1,Q \\
\tOmega^2 &=& \tm ,e \tm ,Q \tm , e Q \tm  \text{.}
\end{eqnarray}
Again, the last two fusion rules are partially undetermined for now, but now have a different structure because $\Omega^2 = m$ in $3D$. The statistics are
\begin{eqnarray}
\theta_e &=& \theta_Q = \theta_{e,Q} = \theta_{Q, \tm} = 0 \\
\theta_{e,\tm} &=& \theta_{Q, \tOmega} = \pi \\
\theta_{e,\tOmega} &=& \pi/2 \\
\theta_{\tm} &=& 0, \pi/2 = \text{?} \\
\theta_{\tOmega} &=&  \text{?} \\
\theta_{\tm,\tOmega} &=& \text{?} \text{.}
\end{eqnarray}
Again, $\theta_{\tm}$ is the same for all basal planes.

We see that the properties of the dimensionally reduced theory can be specified by four pieces of information.  =(There are also fusion rules to specify, but these are determined by the braiding statistics.) First, we can either have $e0$ or $eC$.  Then the three remaining pieces of information are $\theta_{\tm}$, $\theta_{\tOmega}$ and $\theta_{\tm, \tOmega}$.  We will see that the last of these can always be chosen as either $0$ or $\pi/2$, which correspond respectively to integer and half-integer $Z_2^s$ charge of $\tm$.  Following the same notation for $e$, we denote these two possibilities using $0$ and $C$, respectively.  We organize the four pieces of information into the 4-tuple $(e 0/C, \tm 0/C, \theta_{\tm}, \theta_{\tOmega})$.  For example, we would write $(C,0,\pi/2,-\pi/8)$ to describe a dimensionally reduced theory where $e$ has half-charge, $\tm$ has integer charge and statistics $\theta_{\tm} = \pi/2$, and $\theta_{\tOmega} = -\pi/8$.

A very important issue is which redefinitions of the various particles are allowed.  The symmetry charge $Q$ is fixed and cannot be redefined.  The gauge charge $e$ is also fixed by our conventional choice of $\Phi_{e, \Omega} = \theta_{e, \tilde{\Omega}}$. The following redefinitions are allowed:
\begin{eqnarray}
\tOmega &\to& e \tOmega \\
\tOmega &\to& Q \tOmega  \\
\tm &\to& Q \tm \text{.}
\end{eqnarray}
These redefinitions do not affect the fusion rules and statistics angles that we have already fully determined.
Looking at the above redefinitions from the point of view of the $3D$ theory, they all involve binding particles to quasi-strings, and therefore can be done differently for different basal planes.  We thus refer to these as \emph{local} redefinitions. The redefinition $\tm \to e \tm$ is \emph{not} allowed, because we have already made the conventional choice $\theta_{\tm} = \theta_{\tm_I} = \theta_{\tm_m} = 0, \pi/2$.

The following redefinition is also allowed:
\begin{equation}
\tOmega \to \tm \tOmega \text{ and } e \to Q e \text{.}  \label{eqn:Omega-redef}
\end{equation}
However, this redefinition has to be made the same way for all basal planes, because it involves binding together two quasi-strings; that is, it actually is associated with a redefinition $\Omega \to m \Omega$ in the $3D$ theory.  Therefore we refer to it as a \emph{global} redefinition.  At the same time we redefine $\tOmega$, we also have to send $e \to Q e$, in order to keep $\theta_{e,\tOmega}$ fixed.  This redefinition also leaves invariant all the fusion rules and statistics angles that have already been fully determined.

To proceed, we now classify the different possible dimensionally reduced theories. We do this only up to local redefinitions, taking the global redefinition Eq.~(\ref{eqn:Omega-redef}) into account later. We first consider the $e0$ case, then move on to $eC$.

\subsection{Dimensionally reduced SET phases with $e0$}

Recall that for $e0$, we have the following fusion rules:
\begin{eqnarray}
e^2 &=& 1 \\
Q^2 &=& 1 \\
\tm^2 &=& 1,Q \\
\tOmega^2 &=& 1,e,Q,eQ \text{.}
\end{eqnarray}
The statistics are given by
\begin{eqnarray}
\theta_e &=& \theta_Q = \theta_{e,Q} = \theta_{Q, \tm} = 0 \\
\theta_{e,\tm} &=& \theta_{Q, \tOmega} = \pi \\
\theta_{e,\tOmega} &=& 0 \\
\theta_{\tm} &=& 0, \pi/2 = \text{?} \\
\theta_{\tOmega} &=&  \text{?} \\
\theta_{\tm,\tOmega} &=& \text{?} \text{.}
\end{eqnarray}


$\tm^2 = 1$ corresponds to $\tm 0$, and $\tm^2 = Q$ corresponds to $\tm C$.  If $\tm^2 = 1$, we have 
\begin{equation}
0 = \theta_{\tm^2,\tOmega} = 2 \theta_{\tm,\tOmega} \text{,}
\end{equation}
which implies $\theta_{\tm,\tOmega} = 0,\pi$.  We can then redefine $\tm \to Q \tm$ as needed to set $\theta_{\tm, \tOmega} = 0$.  Similarly, if $\tm^2 = Q$, we have
\begin{equation}
\pi = \theta_{\tm^2,\tOmega} = 2 \theta_{\tm,\tOmega} \text{,}
\end{equation}
implying $\theta_{\tm, \tOmega} = \pi/2, 3 \pi/2$, and we can redefine $\tm \to Q \tm$ as needed to set $\theta_{\tm, \tOmega} = \pi/2$.  We have thus fixed $\theta_{\tm, \tOmega}$ (depending on the $\tm^2$ fusion rule), and in doing so we have used up our freedom to redefine $\tm$.

Now, we consider the remaining undetermined information, treating the $\tm 0$ and $\tm C$ cases in turn.  

{\bf Dimensionally reduced theories with $\mathbf{e0 \tm 0}$.} First we take the $\tm 0$ case, considering $2D$ theories with data $(0,0,\theta_{\tm},\theta_{\tOmega})$.  We have
\begin{equation}
\theta_{\tOmega^2, \tm} = 2 \theta_{\tm, \tOmega} = 0 \text{,}
\end{equation}
which implies that only $\tOmega^2 = 1$ or $\tOmega^2 = Q$ are consistent fusion rules.  Also, since $\tOmega^2$ is a boson, we have $\theta_{\tOmega} = \pi q / 2$ for $q = 0,\dots,3$.  Therefore $\theta_{\tOmega,\tOmega} = 2 \theta_{\tOmega} = \pi q$, so $\theta_{\tOmega^2, \tOmega} = 0$.  But this means only
\begin{equation}
\tOmega^2 = 1
\end{equation}
is a consistent fusion rule.  The fusion rules are now completely fixed in this case.

We can now redefine $\tOmega \to Q \tOmega$ as needed to shift $\theta_{\tOmega} \to \theta_{\tOmega} + \pi$, which does not affect any of the fusion rules or other statistics angles. This allows us to choose
\begin{equation}
\theta_{\tOmega} = 0, \pi/2 \text{.}
\end{equation}
The only local redefinition left is $\tOmega \to e \tOmega$.  If we do this alone it modifies $\theta_{\tm, \tOmega}$, so at the same time we can redefine $\tm \to Q \tm$.  The resulting redefinition does not affect any of the fusion rules or statistics angles.  Therefore we have found four possibilities, labeled by $(0,0,\theta_{\tm}, \theta_{\tOmega})$, where $\theta_{\tm}, \theta_{\tOmega}  = 0, \pi/2$.

{\bf Dimensionally reduced theories with $\mathbf{e0 \tm C}$.} Next we take the $\tm C$ case, considering $2D$ theories with data $(0,C,\theta_{\tm},\theta_{\tOmega})$.   We have
\begin{equation}
\theta_{\tOmega^2, \tm} = 2 \theta_{\tm, \tOmega} = \pi \text{,}
\end{equation}
which implies that only $\tOmega^2 = e$ or $\tOmega^2 = e Q$ are consistent fusion rules. 
 
Again $\tOmega^2$ is a boson, so $\theta_{\tOmega} = \pi q / 2$ for $q = 0,\dots,3$, and  $\theta_{\tOmega^2, \tOmega} = 0$.  Therefore we have
\begin{equation}
\tOmega^2 = e \text{,}
\end{equation}
and the fusion rules are fixed. 

Again, we redefine $\tOmega \to Q \tOmega$ as needed to shift $\theta_{\tOmega} \to \theta_{\tOmega} + \pi$, which does not affect any of the fusion rules or other statistics angles. This allows us to choose
\begin{equation}
\theta_{\tOmega} = 0, \pi/2 \text{.}
\end{equation}
Again, the only local redefinition left is $\tOmega \to e \tOmega$.  If we do this alone it modifies $\theta_{\tm, \tOmega}$, so at the same time we can redefine $\tm \to Q \tm$.  The resulting redefinition does not affect any of the fusion rules or statistics angles.  We have thus found four dimensionally reduced theories, labeled by $(0,C,\theta_{\tm}, \theta_{\tOmega})$, where $\theta_{\tm}, \theta_{\tOmega}  = 0, \pi/2$.

{\bf Behavior under global redefinitions.} In total, then, there are eight distinct dimensionally reduced theories with $e0$, with data $(0, 0/C, \theta_{\tm}, \theta_{\tOmega})$, where $\theta_{\tm}, \theta_{\tOmega}  = 0, \pi/2$. These theories are distinct under local redefinitions.  We now would like to consider how they map into one another under the global redefinition Eq.~(\ref{eqn:Omega-redef}).

 $\tOmega \to \tOmega' =  \tm \tOmega$, $e \to e' =  Q e$, which must be made in the same way for all basal planes.

For the $e0 \tm 0$ theories, we find under the global redefinition that $\theta'_{\tOmega} = \theta_{\tOmega} + \theta_{\tm}$, while the other statistics angles, and all the fusion rules, are invariant.  We can thus write
\begin{equation}
(0,0,\theta_{\tm}, \theta_{\tOmega}) \to (0,0,\theta_{\tm}, \theta_{\tOmega} + \theta_{\tm}) \text{.}
\end{equation}
Note that $\theta_{\tOmega} + \theta_{\tm}$ can be chosen to be $0, \pi/2$, as it can be shifted by $\pi$ if needed by making another redefinition $\tOmega \to Q \tOmega$.  So the two $e0 \tm 0$ theories with $\theta_{\tm} = 0$ are invariant under the global redefinition, while the other two, with $\theta_{\tm} = \pi/2$, are exchanged under the redefinition.

For the $e0 \tm C$ theories, we find under the global redefinition that $\theta'_{\tOmega} = \theta_{\tOmega} + \theta_{\tm} + \pi/2$, while the other statistics angles, and all the fusion rules, are invariant.  We can thus write
\begin{equation}
(0,C,\theta_{\tm}, \theta_{\tOmega}) \to (0,C,\theta_{\tm}, \theta_{\tOmega} + \theta_{\tm} + \pi/2) \text{.}
\end{equation}
Similar to the case above, the two $e0 \tm C$ theories with $\theta_{\tm} = \pi/2$ are left invariant, while the two theories with $\theta_{\tm} = 0$ are exchanged.

\subsection{Dimensionally reduced SET phases with $eC$}

Recall that in the $eC$ case the fusion rules are
\begin{eqnarray}
e^2 &=& Q \\
Q^2 &=& 1 \\
\tm^2 &=& 1,Q \\
\tOmega^2 &=& \tm ,e \tm ,Q \tm , e Q \tm  \text{.}
\end{eqnarray}
The statistics are
\begin{eqnarray}
\theta_e &=& \theta_Q = \theta_{e,Q} = \theta_{Q, \tm} = 0 \\
\theta_{e,\tm} &=& \theta_{Q, \tOmega} = \pi \\
\theta_{e,\tOmega} &=& \pi/2 \\
\theta_{\tm} &=& 0, \pi/2 = \text{?} \\
\theta_{\tOmega} &=&  \text{?} \\
\theta_{\tm,\tOmega} &=& \text{?} \text{,}
\end{eqnarray}
with $\theta_{\tm}$ the same for all basal planes.

As before in the $e0$ case, we can use the $\tm \to Q \tm$ redefinition to set $\theta_{\tm,\tOmega} = 0$ when $\tm^2 =1$, and $\theta_{\tm,\tOmega} = \pi/2$ when $\tm^2 = Q$.

{\bf Dimensionally reduced theories with $\mathbf{eC \tm 0}$.} First we consider $\tm^2 = 1$, or $\tm 0$.  In this case we have
\begin{equation}
\theta_{\tOmega^2, \tm} = 2 \theta_{\tOmega, \tm} = 0 \text{.}
\end{equation}
If $\theta_{\tm} = 0$, this implies $\tOmega^2 = \tm, Q \tm$.  On the other hand, if $\theta_{\tm} = \pi/2$, this implies $\tOmega^2 = e \tm, e Q \tm$.

First we consider the case $\theta_{\tm} = 0$.  Then because $\tOmega^2 = \tm, Q \tm$, $\tOmega^2$ is a boson, so $\theta_{\tOmega} = \pi q / 2$ for $q = 0,\dots,3$.  Then
\begin{equation}
\theta_{\tOmega^2, \tOmega} = 2 \theta_{\tOmega,\tOmega} = 4 \theta_{\tOmega} = 0 \text{,}
\end{equation}
and we must have
\begin{equation}
\tOmega^2 = \tm \text{.}
\end{equation}

Now, we consider the redefinition $\tOmega \to \tOmega' =  e \tOmega$, combined with $\tm \to \tm' = Q \tm$.  This redefinition shifts $\theta_{\tOmega} \to \theta_{\tOmega} + \pi/2$, while preserving all other statistics angles and all the fusion rules.  Therefore we can use this to set $\theta_{\tOmega} = 0$.  The remaining local redefinition, $\tOmega \to Q \tOmega$, only shifts $\theta_{\tOmega} \to \theta_{\tOmega} + \pi$, and is thus superfluous.  We have thus found a $2D$ theory with data $(C,0,0,0)$.

Second, we need to consider the case $\theta_{\tm} = \pi/2$.  Then, because $\tOmega^2 = e \tm, e Q \tm$, we have
\begin{equation}
\theta_{\tOmega^2} = \frac{3\pi}{2} = - \frac{\pi}{2} \text{.}
\end{equation}
Therefore, we have for the statistics of $\tOmega$, 
\begin{equation}
\theta_{\tOmega} = \frac{\pi}{2} q - \frac{\pi}{8} , \qquad q = 0,\dots,3 \text{.}
\end{equation}
Then we have
\begin{equation}
\theta_{\tOmega^2, \tOmega} = 2 \theta_{\tOmega,\tOmega} = 4 \theta_{\tOmega} = - \frac{\pi}{2} \text{.}
\end{equation}
This implies we must have the fusion rule
\begin{equation}
\tOmega^2 = e Q \tm \text{.}
\end{equation}

Once again, we consider $\tOmega \to \tOmega' =  e \tOmega$, combined with $\tm \to \tm' = Q \tm$.  Again this shifts $\theta_{\tOmega} \to \theta_{\tOmega} + \pi/2$, while preserving all other statistics angles and all the fusion rules, and we can set  $\theta_{\tOmega} = - \pi / 8$.  The remaining local redefinition, $\tOmega \to Q \tOmega$, only shifts $\theta_{\tOmega} \to \theta_{\tOmega} + \pi$, and is thus superfluous.  We have thus found a $2D$ theory with data $(C,0,\pi/2,-\pi/8)$.

{\bf Dimensionally reduced theories with $\mathbf{eC \tm C}$.}  Now we consider $\tm^2 = Q$, or $\tm C$.  In this case we have
\begin{equation}
\theta_{\tOmega^2, \tm} = 2 \theta_{\tOmega, \tm} = \pi \text{.}
\end{equation}
If $\theta_{\tm} = 0$, this implies $\tOmega^2 = e \tm, e Q \tm$.  On the other hand, if $\theta_{\tm} = \pi/2$, this implies $\tOmega^2 =  \tm,  Q \tm$.

First we consider the case $\theta_{\tm} = 0$.  Then because $\tOmega^2 = e \tm, e Q \tm$, we have $\theta_{\tOmega^2} = \pi$, implying
\begin{equation}
\theta_{\tOmega} = \frac{\pi}{2} q + \frac{\pi}{4} , \qquad q = 0,\dots,3 \text{.}
\end{equation}
Then we have
\begin{equation}
\theta_{\tOmega^2, \tOmega} = 2 \theta_{\tOmega,\tOmega} = 4 \theta_{\tOmega} = \pi \text{,}
\end{equation}
which implies we have the fusion rule
\begin{equation}
\tOmega^2 = e \tm \text{.}
\end{equation}

Again, we consider the redefinition $\tOmega \to \tOmega' =  e \tOmega$, combined with $\tm \to \tm' = Q \tm$.  Again this shifts $\theta_{\tOmega} \to \theta_{\tOmega} + \pi/2$, while preserving all other statistics angles and all the fusion rules, and we can set  $\theta_{\tOmega} = \pi / 4$.  The remaining local redefinition, $\tOmega \to Q \tOmega$, only shifts $\theta_{\tOmega} \to \theta_{\tOmega} + \pi$, and is thus superfluous.  We have thus found a $2D$ theory with data $(C,C,0,\pi/4)$.

Second, we need to consider the case $\theta_{\tm} = \pi/2$.  Then because $\tOmega^2 = \tm, Q \tm$, we have $\theta_{\tOmega^2} = \pi/2$, implying
\begin{equation}
\theta_{\tOmega} = \frac{\pi}{2} q + \frac{\pi}{8} , \qquad q = 0,\dots,3 \text{.}
\end{equation}
Then we have
\begin{equation}
\theta_{\tOmega^2, \tOmega} = 2 \theta_{\tOmega,\tOmega} = 4 \theta_{\tOmega} = \frac{\pi}{2} \text{,}
\end{equation}
which implies we have the fusion rule
\begin{equation}
\tOmega^2 = \tm \text{.}
\end{equation}

Again, we consider the redefinition $\tOmega \to \tOmega' =  e \tOmega$, combined with $\tm \to \tm' = Q \tm$.  Again this shifts $\theta_{\tOmega} \to \theta_{\tOmega} + \pi/2$, while preserving all other statistics angles and all the fusion rules, and we can set  $\theta_{\tOmega} = \pi / 8$.  The remaining local redefinition, $\tOmega \to Q \tOmega$, only shifts $\theta_{\tOmega} \to \theta_{\tOmega} + \pi$, and is thus superfluous.  We have thus found a $2D$ theory with data $(C,C,\pi/2,\pi/8)$.

{\bf Behavior under global redefinitions.}  In total, there are four distinct $2D$ theories with $eC$, with data $(C,0,0,0)$, $(C,0,\pi/2,-\pi/8)$, $(C,C,0,\pi/4)$, and $(C,C,\pi/2,\pi/8)$.  Let's consider the behavior of these under the global redefinition $\tOmega \to \tOmega' = \tm \tOmega$.  In order to preserve the parameter $\theta_{e, \tOmega}$, we must also redefine $e \to e' = Q e$.  In addition, if $\theta_{\tm} = \pi/2$, to preserve $\theta_{\tm, \tOmega}$, we must redefine $\tm \to \tm' = Q \tm$. We find that all four $2D$ theories are invariant under the global redefinition.

\section*{Appendix C: Fractionalization patterns}

Here, we use the results of Appendix B to describe the possible fractionalization patterns in $Z_2Z_2$ SET phases.  The description here is equivalent to that given in the main text, but is more detailed in accounting for all the properties of the $2D$ SET orders upon dimensional reduction.

Obviously, one piece of information in the fractionalization pattern is the symmetry charge of $e$, so we have either $e0$ or $eC$.  The remaining information has to do with symmetry action on $m$ quasi-strings.  Using the dimensional reduction approach, a crucial point is that the symmetry action on $m$ is encoded in differences of the properties from one basal plane to another.  Because $\theta_{\tm}$ is the same in all basal planes, we then have two pieces of information.  One is written $m_m 0 / C$, which expresses whether the difference in symmetry charge of $\tm$ between the two basal planes is integer or fractional.  The other is written $\Omega_m 0 / s$, which has to do with the difference $\theta^{[m]}_{\tOmega} - \theta^{[I]}_{\tOmega}$.  If this difference is $0 \operatorname{mod} \pi$, we write $\Omega_m 0$, and if it is $ \pi/2 \operatorname{mod} \pi$, we write $\Omega_m s$, where the ``$s$'' stands for semion.  Putting this information together, we would write \emph{e.g.} $e0m_m 0 \Omega_m 0$ to specify the entire fractionalization pattern.

Each fractionalization pattern corresponds to several different choices of dimensionally reduced theories on the $I$ and $m$ basal planes.  Therefore, it is important to be sure that the information given in the fractionalization pattern is always well-defined under global redefinitions, for all possible choices of dimensionally reduced theories.  Or, if some information is not well-defined, it should be ill-defined for \emph{all} possible choices of dimensionally reduced theories corresponding to a given fractionalization pattern.

Let's first check this for the $e0 m_m 0 \Omega_m 0$ fractionalization pattern.  We first note that it does not matter which dimensionally reduced theory occurs on which basal plane.  For $e0$, the global redefinition does not affect the charge of $\tm$, so $m_m 0$ is certainly well-defined.  However, the statistics of $\tOmega$ can change under global redefinition, so we have to be more careful.  One possibility is that both basal planes have the same theory with data $(0,0,\theta_{\tm},\theta_{\tOmega})$.  While this data changes under global redefinition if $\theta_{\tm} = \pi/2$, the difference in $\theta_{\tOmega}$ between the two planes is unchanged, and $\Omega_m 0$ is thus well-defined.  The other possibility  is that both basal planes have the same theory with data $(0,C,\theta_{\tm},\theta_{\tOmega})$, where again the difference of $\theta_{\tOmega}$ between the planes is unchanged by global redefinition.

Next, we consider $e0 m_m 0 \Omega_m s$.  Again, $m_m 0$ is well-defined.  Here, one possibility is that one basal plane has data $(0,0,\theta_{\tm}, \theta_{\tOmega})$, while the other has data $(0,0,\theta_{\tm}, \theta_{\tOmega} + \pi/2)$.  The difference in $\theta_{\tOmega}$ between the planes is unchanged under global redefinition.  This is also clearly true for the other possibility, which is that one plane has data $(0,C,\theta_{\tm}, \theta_{\tOmega})$, while the other has data $(0,C,\theta_{\tm}, \theta_{\tOmega}+\pi/2)$.

Next, we consider $e0 m_m C$.  We do not specify $\Omega_m$ here, because we will see it is not well-defined.  The most general possibility is that one plane has data $(0,0,\theta_{\tm}, \theta^{[a]}_{\tOmega} )$, while the other has  $(0,C,\theta_{\tm}, \theta^{[b]}_{\tOmega})$.  The difference $m_m C$ is clearly well-defined.  Under global redefinition we have
\begin{eqnarray}
(0,0,\theta_{\tm}, \theta^{[a]}_{\tOmega} ) &\to& (0,0,\theta_{\tm}, \theta^{[a]}_{\tOmega}  + \theta_{\tm}) \\
(0,C,\theta_{\tm}, \theta^{[b]}_{\tOmega}) &\to& (0,C,\theta_{\tm}, \theta^{[b]}_{\tOmega} + \theta_{\tm} + \pi/2) \text{.}
\end{eqnarray}
Therefore, the difference $\theta^{[b]}_{\tOmega} - \theta^{[a]}_{\tOmega}$ shifts by $\pi/2$ under global redefinition, and is thus not well-defined.  (Note that it is also consistently ill-defined, for all possible choices of dimensionally reduced theories.)

Now we consider fractionalization patterns with $eC$.  In this case, for a fixed $\theta_{\tm}$, there are only two possible dimensionally reduced theories, which have $\tm 0$ and $\tm C$. Global redefinition leaves all these $2D$ theories invariant, so it plays no role.

First we consider the $eC m_m 0$ fractionalization pattern.  In this case, both $I$ and $m$ basal planes have to be in the same dimensionally reduced theory -- any one of the four choices is fine.  The statistics of $\tOmega$ is thus the same in both basal planes, so we don't need to specify anything about $\Omega_m$.

Finally we consider $eC m_m C$.  In this case, for a given $\theta_{\tm}$, the $I$ basal plane is in one of the two possible theories, and the $m$ plane is in the other.  We see that the difference $\theta^{[m]}_{\tOmega} - \theta^{[I]}_{\tOmega} = \pm \pi/4$.  At first glance, it looks like there might be two possibilities here, having to do with the plus or minus sign, but $\pm \pi/4$ are equivalent under local redefinitions that shift $\theta_{\tOmega} \to \theta_{\tOmega} + \pi/2$ in one of the basal planes.  Therefore there is a single $eC \Omega_m C$ fractionalization pattern.

\end{document}